\documentclass[11pt,a4paper]{article}
\pdfoutput=1
\usepackage{jheppub}
\usepackage{amsmath,amsfonts,amsthm,graphicx,ulem}
\usepackage{float}

\newcommand{\comment}[1]{}
\newcommand{\bal}{\begin{align}}
\newcommand{\eal}{\end{align}}
\newcommand{\beq}{\begin{equation}}
\newcommand{\eeq}{\end{equation}}
\newcommand\beqa{\begin{eqnarray}}
\newcommand\eeqa{\end{eqnarray}}
\newcommand\bea{\begin{array}}
\newcommand\eea{\end{array}}

    \newcommand{\COMMENT}[1]{}
    
    \newcommand{\neqa}{\nonumber\end{eqnarray}}

\def\o{{\omega}}

\def\[{\left[}
\def\]{\right]}

\def\s{\sigma}

\def\b{\beta}

\def\D{\Delta}

\def\<{\langle}
\def\>{\rangle}

\def\tr{\text{tr}~}
\def\Tr{\text{Tr}~}
\def\Ub{{\bf U}}
\def\Vb{{\bf V}}

\def\i2{\frac{i}{2}}

\title{ \center{ Two-point correlator of twist-2 light-ray operators in N=4 SYM in BFKL approximation}}
\author[a,b]{Ian Balitsky}
\author[c,d,1]{Vladimir Kazakov%
\note{member of Institut Universitaire de France}}
\author[c]{Evgeny Sobko}

\affiliation[a]{Physics Dept., Old Dominion University, Norfolk VA 23529}
\affiliation[b]{Theory Group, JLAB, 12000 Jefferson Ave, Newport News, VA 23606}
\affiliation[c]{Ecole Normale Superieure, LPT, 24 rue Lhomond,  75231 Paris CEDEX-5,
  France}
\affiliation[d]{Universit\'e Pierre et Marie Curie, Paris-VI, France}

\emailAdd{balitsky AT jlab.org}
\emailAdd{evgenysobko AT gmail.com}
\emailAdd{kazakov AT lpt.ens.fr}

\abstract{We generalize local operators of the leading twist-2 of N=4 SYM theory to the case of  complex Lorentz spin \(j\) using  principal series representation of \(sl(2,R)\). We give the direct computation of correlation function of two such non-local operators  in the BFKL regime when \(j\rightarrow 1\).  The correlator appears to have the expected conformal coordinate
  dependence governed by the anomalous dimension of twist-2 operator in NLO BFKL approximation predicted by Kotikov and Lipatov.}

\keywords{N=4 SYM, NLO BFKL, Twist-2, Wilson loops, Correlator, principal series}
\subheader{LPT ENS-13/20 }

\usepackage{varioref}
\usepackage{makeidx}
\makeindex
\usepackage{amssymb}

\begin{document}

  \maketitle

\section{Introduction}

Quantum integrability of the planar N=4 SYM theory \cite{Beisert:2010jr} gave hopes for a rather complete understanding of the full  dynamics of this superconformal 4D theory, as well as for providing us with  efficient methods of computation of the  basic physical quantities: spectrum of anomalous dimensions,  correlators of local and non-local operators, amplitudes. But the actual computations are still very involved and usually relay on various approximations, such as weak or strong coupling, or the BFKL limit.  For the N=4 SYM spectral problem,    there has been a lot of progress  in the last years  \cite{Beisert:2010jr}  allowing to study it not only at these approximations but also numerically, at any coupling.  Recently these developments have been culminated in the formulation of a well defined system of Riemann-Hilbert equations  \cite{Gromov:2013pga}.  But for the correlation functions the situation is far more complicated and one is here on the early stage of case-by-case study in a weak \cite{Okuyama:2004bd,Roiban:2004va,Alday:2005nd,Plefka:2012rd,Eden:2011we,Eden:2012rr,Gromov:2012vu,Kostov:2012yq,Kazakov:2012ar,Alday:2013cwa} or strong \cite{Costa:2010rz,Zarembo:2010rr,Janik:2011bd,Kazama:2012is,Buchbinder:2011jr,Russo:2010bt} coupling regime.

In these circumstances, the   BFKL approximation \cite{Kuraev:1976ge,Kuraev:1977fs,Balitsky:1978ic} appears to be an interesting testing ground for the understanding of general properties of N=4 SYM theory.  The BFKL approximation was originally proposed for the study of the Regge (collinear) limit of hadron deep inelastic scattering amplitudes in QCD, when \(g^2_{YM}N\to 0\), the Mandelstam variable \(s\to\infty\) with  \(g^2_{YM}N\log \frac{s}{M^2}\) - fixed. Kotikov and Lipatov  \cite{Kotikov:2002ab}  applied a similar approximation to the study of  anomalous dimensions of twist-2 operators \(\tr [F^{+\mu}_{\ \ \bot}g^{\bot}_{\mu\nu}D^{j-2}F^{\nu +}_{\bot}+...]\) in the BFKL limit, when the Lorentz spin is analytically continued to \( j-1=\o\to 0\), the 't~Hooft coupling \(g^2=\frac{g^2_{{}_{YM}}N}{16\pi^2}\to 0\) and  \(\frac{g^2}{j-1}=const\).   Their explicit formula has passed a few tests, and  in particular it was confirmed by the 4-loop  \cite{Bajnok:2008qj} and then 5-loop computation from the L\"uscher formula \cite{Lukowski:2009ce}. Also significant progress was made in the strong-coupling regime \cite{Brower:2006ea,Costa:2012cb,Kotikov:2013xu}

But so far, to our knowledge,   no reliable  definition was given for the analytic continuation of these operators to a complex Lorentz spin \(j\). Usually  one performs this continuation in the final results for dimensions, without elaborating on the explicit definition of the generalized operator. The   principle of maximal transcendentality, commonly used for the analytic continuation from the integer spins  within each order of the perturbation theory,  was never proven\footnote{ In the \cite{Janik:2013nqa}  a certain analyticity condition for the Baxter Q-function of \(sl(2)\) Heisenberg spin chain were proposed  reproducing the analytic continuations of harmonic sums w.r.t. the spin, at one and two loops.}. The goal of this paper is to construct an explicit form of twist-2 operators for arbitrary  complex Lorentz spin and to perform the direct calculation of their two-point correlation function in the Leading Logarithmic Approximation (LLA) BFKL with Leading Order (LO) accuracy for the impact factors and NLO for the anomalous dimension.
 It can be considered as a necessary step for a slightly more ambitious goal - the computation of 3-point correlators
and  the corresponding OPE structure functions, in the BFKL approximation.

Let us describe the logic of our approach. We start with construction of a non-local light-ray twist-2 operator which transforms according to the principal series representation of \(sl(2,R)\) with conformal spin \(J=\frac{1}{2}+i\nu,\ \nu\in\mathbb{R}\) and even parity. This operator   diagonalizes the renormalization group Hamiltonian.  It is constructed from two local fields, with the coordinates \(x_{1-}\) and \(x_{2-}\) on the same light-ray, connected by the adjoint Wilson line factor. The operator is then integrated over the positions of both local operators along the light-ray.

\begin{figure}[H]
\center{\includegraphics[scale=0.8]{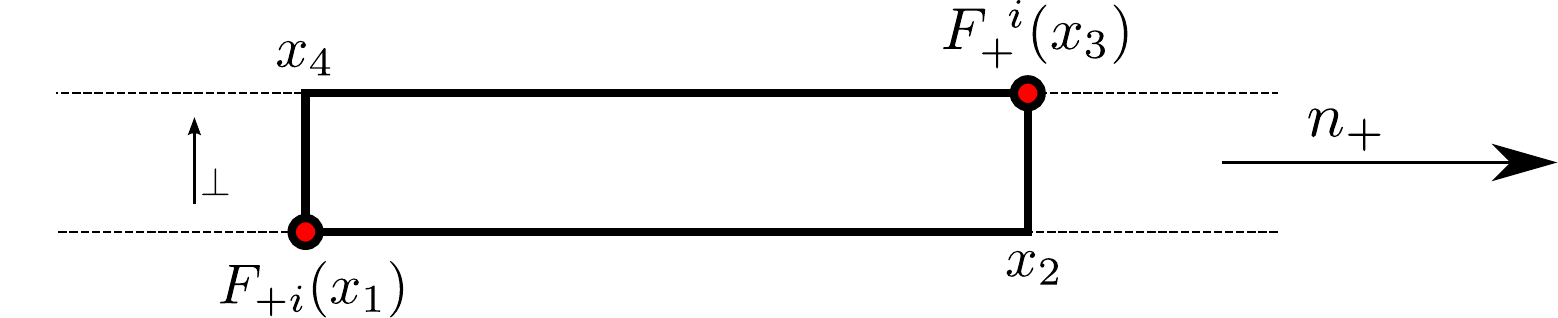} \\}
\caption{The ``frame" configuration for the regularized  light ray operator: long sides are stretched along  the light ray with the direction \(n_+\), short sides oriented in an orthogonal direction.   }
\label{ris:Ramka}
\end{figure}

Constructed in this way, the light-ray operator is   a singular, not well defined object in the  BFKL regime.  To avoid the singularity, we regularize it  by placing  the local operators on two diffferent, but very close parallel light rays separated by a small distance   \(\delta x_\bot=|x_{1\bot}-x_{3\bot}|\)    from each other, in a direction orthogonal to the light rays (see Fig.\ref{ris:Ramka}). We close it into a rectangular  Wilson loop, with two fields inserted at the diagonally opposite corners of the loop, as depicted in Fig.\ref{ris:Ramka}.
We will call this loop the Wilson frame.\footnote{we hope  the reader will avoid the confusion between this Wilson frame and the coordinate frame}
Note that  under a generic conformal transformation the frame will look almost the same: the distance  between two light-lines will be slightly changed and the short lines connecting the ends of light ray intervals will be only slightly deformed. One can show that this deformation does not change the final results in our approximation: one can neglect the gluons emitted by infinitesimally short sides of the frame.

\begin{figure}[H]
\center{\includegraphics[scale=0.6]{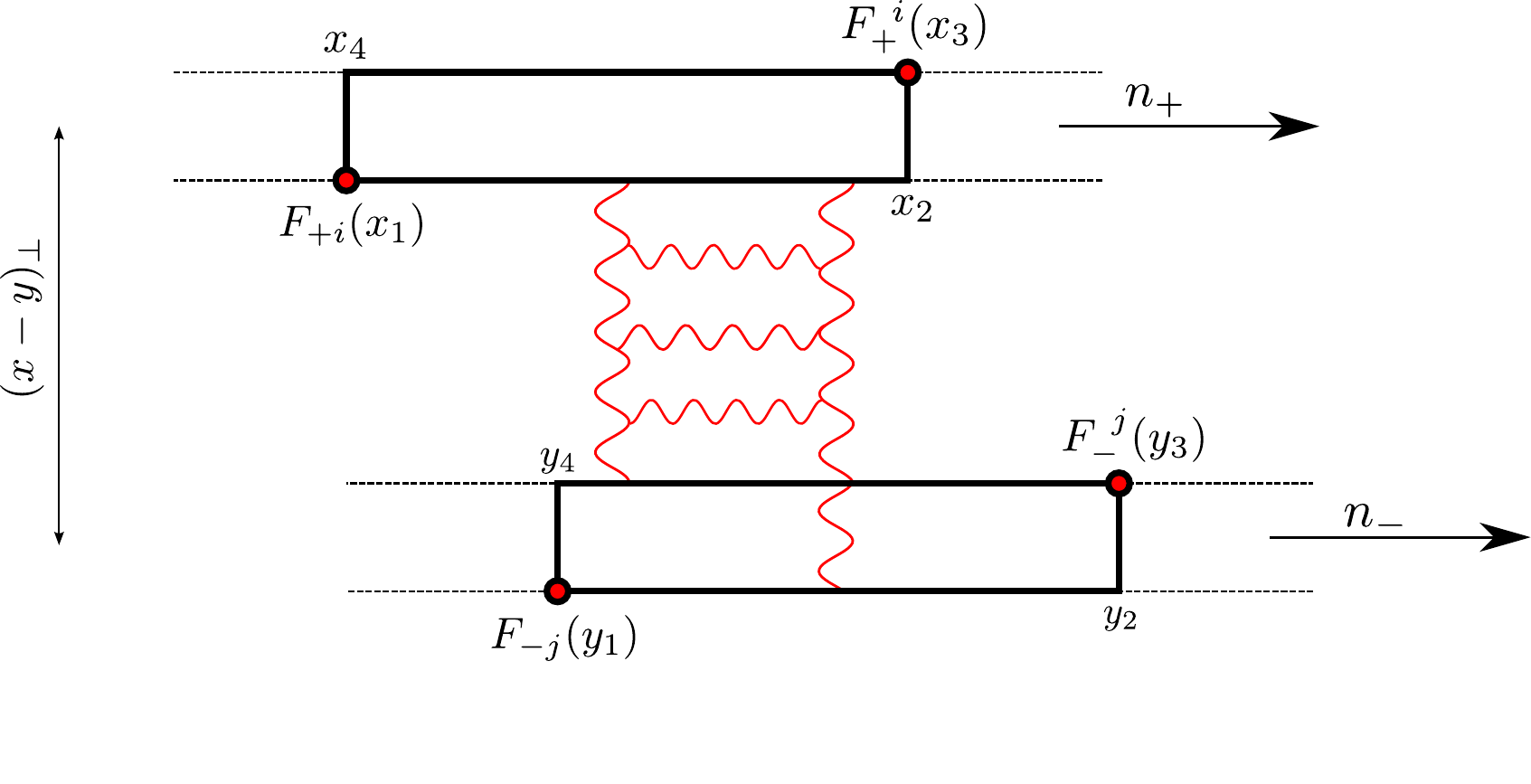} \\}
\caption{Two Wilson frames, at a distance \(|x-y|_\bot\) from each other, stretched in two  different light-cone directions \(n_+\) and \(n_-\) and a typical gluon exchange between them.  }
\label{ris:2Ramki}
\end{figure}

 We will calculate in this paper   the   correlation function of two such   objects separated by a certain distance in orthogonal space and  stretched along two different  light-like directions given by vectors  \(n_+\) and \(n_-\), as shown in Fig.\ref{ris:2Ramki}. We will use for that the  OPE\footnote{For the recent development of these ideas see \cite{Caron-Huot:2013fea}.Another type of OPE for Wilson Loops with null edges was elaborated in \cite{Alday:2010ku,Gaiotto:2010fk,Sever:2011da,Basso:2013vsa,Basso:2013aha}} decomposition over ``colour dipoles" in the limit when \((x_1-x_3)^2\to 0\), proposed by one of the  authors \cite{Balitsky:1995ub}(see also the review \cite{Balitsky:2001gj}). The ``colour dipole"   is a pair of  parallel infinite light-like Wilson lines, with a cut-off \(\sigma\) on the momenta of  gauge field in the light-cone  direction. After such decomposition,  symbolically depicted in Fig.\ref{ris:Factorisation}, we  calculate the correlator between two colour dipoles. This calculation is done in two steps: first,  for each dipole we compute the correlator for small values of the cutoff \(\tilde{\sigma}\), such that \(g^2\log\frac{\tilde{\sigma}}{\sigma_0}\ll 1\), where \(\sigma_0\ < \tilde{\sigma}\ll \sigma\), when the lowest order of perturbation theory dominates in the LLA approximation, and then we evolve the result w.r.t. \(\tilde{\sigma}\) to its final value \(\sigma\). It is important to stress that the evolution with respect to the scale \(\sigma_{\pm}\)  for each colour dipole is governed by the BFKL equation \cite{Balitsky:1995ub}. The ratio of cut-offs  \(\frac{\s_+\s_-}{\s_{0+}\s_{0-}}\), due to the conformal invariance, appears to be related to certain anharmonic ratios defined by the shapes of our configuration of frames. The last step is the integration over the coordinates of Wilson frame along each of the light-rays.
In what follows, we are going to precise each step of this calculation.

\section{Generalisation of twist-2 operators to the case of principal series $sl(2,R)$.}
The twist-2 supermultiplet of local operators was explicitly constructed in \cite{Belitsky:2003sh}. For example, one of the components at zero order in \(g_{{}_{YM}}\)  reads as follows (\(j\) is even):
\begin{equation}
\mathcal{S}_{\rm loc}^{j}(x)=6\mathcal{O}_{gg}^{j}(x)+\frac{j-1}{2}
\mathcal{O}_{qq}^{j}(x)+\frac{j(j-1)}{4}\mathcal{O}_{ss}^{j}(x),\label{LocOper}
\end{equation}
where
\begin{eqnarray}
\mathcal{O}_{gg}^{j}(x)&=& \tr \mathcal{G}^{\frac{5}{2}}_{j-2,x_1,x_2} F^{\ \ \mu}_{+\bot}(x_1)g^{\bot}_{\mu\nu}F^{\ \ \nu}_{+\bot}(x_2)|_{x=x_1=x_2},\\
\mathcal{O}_{qq}^{j}(x)&=& \tr \mathcal{G}^{\frac{3}{2}}_{j-1,x_1,x_2} \bar{\lambda}_{\dot{\alpha}A}\sigma^{+\dot{\alpha}\beta}(x_1)\lambda^A_\beta(x_2)|_{x=x_1=x_2},\\
\mathcal{O}_{ss}^{j}(x)&=& \tr \mathcal{G}^{\frac{1}{2}}_{j,x_1,x_2} \bar{\phi}_{AB}(x_1)\phi^{AB}(x_2)|_{x=x_1=x_2}.
\end{eqnarray}
We   introduced here  the  differential operator \(\mathcal{G}^{\alpha}_{n,x_1,x_2}=i^n( \nabla_{x_2} +\nabla_{x_1})^n C_n^{\alpha}(\frac{\nabla_{x_2}-\nabla_{x_1}}{\nabla_{x_2}+\nabla_{x_1}})\), where \(C_n^{\alpha}(x)\) is the Gegenbauer polynomial of order \(n\) with index \(\alpha\). \(\nabla_x\) are covariant derivatives in
the light-like direction \(n_+\): \(\nabla_x=n_+^\mu(\partial_\mu-ig_{{}_{YM}}A_\mu)=\partial_+-ig_{{}_{YM}}A_+\). The fields entering the operators belong to the set \(X=\{F^{\ \ \mu}_{+ \bot}, \lambda^A_{+\alpha},\bar{\lambda}^{\dot{\alpha}}_{+A},\phi^{AB}\}\) which contains the field components with maximal spin (see appendix \ref{AppNotations}).

Let us note that all  components of twist-2 supermultiplet are constructed from so called primary  conformal operators, in the sense that they realize the highest-weight representation of \(sl(2,R)\). For example, in the  case of \(\mathcal{S}_{\rm loc}^{j}\) the operators \(\mathcal{O}_{gg}^{j},\ \mathcal{O}_{qq}^{j},\ \mathcal{O}_{ss}^{j}\) are primaries, with conformal spin \(J=j+1\). Due to  supersymmetry we should work with superconformal operators transforming under an irreducible representation of \(sl(2|4)\). It leads to the superprimary operators which are a linear combination of conformal operators as in (\ref{LocOper}). It is important to stress that the coefficients in this combination do not depend on the Yang-Mills coupling constant \(g_{{}_{YM}}^2\) and the  renormalization takes place  for each conformal operator separately\footnote{It is so, because the supercharges don't depend on \(g_{{}_{YM}}\)}. These superconformal operators  diagonalize the one-loop dilatation operator given by the Hamiltonian:

\begin{gather}
H=g^2[H_{12}+H_{21}],\\
H_{i,i+1}\phi(z_i,z_{i+1})=2\left[\psi(J^{\mathfrak{G}}_{i,i+1})-\psi(1)\right],
\end{gather}
where \(J^{\mathfrak{G}}_{i,i+1}\) is defined through the Casimir operator \(J_{i,i+1}^2=J^{\mathfrak{G}}_{i,i+1}(J^{\mathfrak{G}}_{i,i+1}-1)\) of the full  \(\mathfrak{G}=PSU(2,2|4)\) group \cite{Beisert:2003yb}.

We start with generalization of local conformal operators. Our logic will be  close to the logic of \cite{Balitsky:1987bk}. Local conformal operators correspond to the discrete unitary irreps of \(sl(2,R)\). Let us construct a nonlocal light-ray operator \footnote{An interesting dual conformal symmetry on the light-cone was discovered in \cite{Derkachov:2013bda}.} which realizes the principal series irrep of \(sl(2,R)\) with the conformal spin \(J=\frac{1}{2}+i\nu,\ \nu\in\mathbb{R}\) and even parity.
A general light-ray operator with local fields \(\chi^s\) of the same conformal spin \(s\) looks as follows:
\begin{gather}
S^s_{\phi}(x_{1\bot})=\int\limits_{-\infty}^{\infty} dx_{1-}\int\limits_{x_{1-}}^{\infty} dx_{2-} \phi(x_{1-},x_{2-})\chi^s(x_{1})[x_{1},x_{2}]_{Adj}\chi^s(x_{2}), \label{non loc tw2 oper}
\end{gather}
where
\begin{equation}
[x_1,x_2]_{Adj}=P\text{exp}[ig_{{}_{YM}}\int\limits_0^1du(x_2-x_1)_\mu A_{Adj}^\mu(x_1(1-u)+x_2u)],
\end{equation}
and the function \(\phi(x_1,x_2)\) is an arbitrary function of two variables. We are looking for the operators \(S^s_{\phi}\) which are the  eigenfunctions of  \(sl(2,R)\)  Casimir
operator defined in the following way:
Take the generators of \(sl(2,R)\) satisfying standard relations:
\begin{gather}
\left[J_3,J_\pm\right]=\pm J_\pm,\notag\\
\left[J_+,J_-\right]=-2 J_3 .
\end{gather}
and  realize them on the fields with conformal spin \(s\):
\begin{gather}
J_+=\frac{i}{\sqrt{2}}P_+=\frac{1}{\sqrt{2}}\frac{d}{dx},\notag\\
J_-=\frac{i}{\sqrt{2}}K_+=\sqrt{2}(2sx+x^2\frac{d}{dx}),\\
J_3=\frac{i}{2}(D+M_{-+})=s+x\frac{d}{dx}.\notag
\end{gather}
Here \(x\) is a coordinate along the light ray.
The equation on the eigenvalues and eigenfunctions \(\phi(x_{1-},x_{2-})\)  of the Casimir operator
\begin{gather}
\vec{J}^2S^s_\phi=J(J-1)S^s_\phi=(j+1)jS^s_\phi
\end{gather}
can thus be rewritten as a partial differential equation
\begin{gather}
\left[\beta^2\left(\frac{\partial^2}{\partial \beta^2}-\frac{\partial^2}{\partial \alpha^2}\right)-2s \beta \frac{\partial}{\partial \beta}+s(s+1)\right]\phi(\alpha,\beta)=J(J-1)\phi(\alpha,\beta),
\end{gather}
where \(\alpha=x_{1-}+x_{2-}, \ \beta=x_{2-}-x_{1-}\). Separating the variables \(\phi(\alpha,\beta)=f(\alpha)g(\beta)\) we get:
\begin{equation}
\left\{
\begin{aligned}
\frac{\partial^2}{\partial \alpha^2}f(\alpha)&=-k^2 f(\alpha),\\
\left(\beta^2\frac{\partial}{\partial \beta^2}-2s\beta\frac{\partial}{\partial \beta}+s(s+1)+k^2\beta^2\right)g(\beta)&=J(J-1)g(\beta).\\
\end{aligned}
\right.
\end{equation}
The general solution for the eigenfunctions reads as follows:
\begin{gather}
\frac{\phi(x_{1-},x_{2-})}{(x_{2-}-x_{1-})^{2s-\frac{3}{2}}}=\notag\\
=\int dk \,\,\left[\eta_1(k)e^{ik(x_{1-}+x_{2-})}(C_{11} {\bf J}_{-\frac{1}{2}+J}(k(x_{2-}-x_{1-}))+C_{12}{\bf J}_{\frac{1}{2}-J}(k(x_{2-}-x_{1-})))\right.+\notag\\
+ \left. \eta_2(k)e^{-ik(x_{1-}+x_{2-})}(C_{21} {\bf J}_{-\frac{1}{2}+J}(k(x_{1-}-x_{2-}))+C_{22}{\bf J}_{\frac{1}{2}-J}(k(x_{2-}-x_{1-}))\right],
\end{gather}
where \({\bf J}_\nu(x)\) - is a Bessel function and \(\eta_1(k),\ \eta_2(k)\) are arbitrary functions of \(k\).
In addition, we should impose a set of constraints on the operator \(S^s_\phi\). First of all, it should be an entire function of \(J\), to allow for an unambiguous  analytic continuation of the light-ray operator in \(J\), and the dimension of this operator should  coincide with the standard local twist-2 operator for any  integer \(J\).  Both of these conditions are satisfied if we choose a linear combination of Bessel functions as the  Hankel function of the second order:
\begin{gather}
C_{i1} {\bf J}_{-\frac{1}{2}+J}(k(x_{2-}-x_{1-}))+C_{i2}{\bf J}_{\frac{1}{2}-J}(k(x_{2-}-x_{1-})) \rightarrow {\bf H}^2_{J-\frac{1}{2}}(k(x_{2-}-x_{1-})), \ \ \ i\in\{1,2 \}
\end{gather}
In this way we obtain an operator which is an entire function of spin: It is well defined for \(J=\frac{1}{2}+i\nu\) and thus it can be uniquely continued to the whole complex plane of \(J\).
It is natural to choose  the so far arbitrary coefficient functions as   \(\eta_1(k)=\eta_2(k)=\frac{1}{2}\delta(k)(\frac{k}{2})^{J-\frac{1}{2}}\) which naturally sets   to zero  the center-of-mass momentum  \(k\)  and cancels  the singularity at \(k\to 0\).

Now, using the asymptotics of Hankel function at \(k\rightarrow 0\):
\begin{gather}
{\bf H}^2_{J-\frac{1}{2}}(k(x_{2-}-x_{1-})) \rightarrow -(\frac{k(x_{2-}-x_{1-})}{2})^{-J+\frac{1}{2}}\frac{\Gamma(J-\frac{1}{2})}{\pi},
\end{gather}
we get the following form of the light-ray operators (denoted by \(\breve S\)) for scalars, fermions and gluons:
\begin{gather}
\breve{S}_{sc}^J(x_{1\bot})=\int\limits_{-\infty}^{\infty} d x_{1-} \int\limits_{x_{1-}}^{\infty}dx_{2-} (x_{2-}-x_{1-})^{-J}\ \tr\bar{\phi}_{AB}(x_{1})[x_1,x_2]_{Adj}\phi^{AB}(x_{2}), \label{nonlocSc}\\
\breve{S}_{f}^J(x_{1\bot})=\int\limits_{-\infty}^{\infty} d x_{1-} \int\limits_{x_{1-}}^{\infty}dx_{2-}  (x_{2-}-x_{1-})^{-J+1}\ \tr\bar{\lambda}_{\dot{\alpha}A}(x_{1})\sigma^{+\dot{\alpha}\beta}[x_1,x_2]_{Adj}\lambda^A_\beta(x_{2}),\\
\breve{S}_{gl}^J(x_{1\bot})=\int\limits_{-\infty}^{\infty} d x_{1-} \int\limits_{x_{1-}}^{\infty}dx_{2-} (x_{2-}-x_{1-})^{-J+2}\ \tr F^{\ \ \mu}_{+\bot}(x_{1})g^{\bot}_{\mu\nu}[x_1,x_2]_{Adj}F^{\ \ \nu}_{+\bot}(x_{2}).\label{nonlocGluon}
\end{gather}
Let us clarify the correspondence of nonlocal operators (\ref{nonlocSc})-(\ref{nonlocGluon}) to the local operators, using gluons as an example. We take an odd  integer \(J\)  in (\ref{nonlocGluon})  and define the integral  over  \(x_2-x_1\) , with the prescription analogous to the eq.(3.19) of \cite{Balitsky:1987bk}.    This gives, e.g. for the gluons:
\begin{gather}
\breve{S}_{gl}^J(x_{1\bot})= \frac{2^{3-J}2\pi i}{\Gamma(J-2)}\int\limits_{-\infty}^{\infty}dx_-\, \tr\left[(\overleftarrow{\nabla}-\overrightarrow{\nabla})^{J-3}
 F^{\ \ i}_{+}(x)F_{+i}(x)\right],\label{NLsc}
\end{gather}
where \(\overrightarrow{\nabla}\) and \(\overleftarrow{\nabla}\) are covariant derivatives which act on the left and right scalars, correspondingly. On the other hand, the local gluon operator (for odd \(J\)) has the following form:
\begin{gather} \mathcal{O}^{j}_{gl}(x)=\mathcal{O}^{J-1}_{gl}(x)=\tr\left[i^{J-3}(\overrightarrow{\nabla}+\overleftarrow{\nabla})^{J-3}C_{J-3}^{\frac{5}{2}}\left(\frac{\overrightarrow{\nabla}-\overleftarrow{\nabla}}{\overrightarrow{\nabla}+\overleftarrow{\nabla}}\right)
F^{\ \ i}_{+}(x)F_{+i}(x)\right].
\end{gather}
Integrating it over the coordinate, we get:
\begin{gather}
\int\limits_{-\infty}^{\infty}dx_-\,\mathcal{O}^{J-1}_{gl}(x)=i^{J-3}\frac{\Gamma(J-\frac{1}{2})2^{J-3}}{\Gamma(\frac{5}{2})\Gamma(J-2)}
\int\limits_{-\infty}^{\infty}dx_-\tr\left[(\overrightarrow{\nabla}-\overleftarrow{\nabla})^{J-3} F^{\ \ i}_{+}(x)F_{+i}(x)\right].\label{Lsc}
\end{gather}
All terms in \(\mathcal{O}^{J-1}_{gl}(x)\) with nonzero power \(\overrightarrow{\nabla}+\overleftarrow{\nabla}\) disappear because they are full derivatives.
Now we want to construct a nonlocal superconformal operator. Nonlocal operator which corresponds to  \(\int dx\, \mathcal{S}_{loc}^{j}(x)\) is a sum \(\mathcal{S}_{nloc}^{J}=c_{sc}S_{sc}^J+c_{f}S_{f}^J+c_{gl}S_{gl}^J\) with some so far unknown coefficients. They can be fixed by comparing the local and nonlocal operators in the case of integer \(J\). For example for gluons, using (\ref{NLsc}), (\ref{Lsc}) we conclude that
\begin{gather}
c_{gl}=\frac{i^J 2^{2J-4}\Gamma(J-\frac{1}{2})}{\pi \Gamma(\frac{1}{2})}.
\end{gather}
And similarly we get the other coefficients:
\begin{gather}
c_f=i\frac{J-2}{2}\frac{i^J 2^{2J-4}\Gamma(J-\frac{1}{2})}{\pi \Gamma(\frac{1}{2})},\\
c_{sc}=-(J-2)(J-1)\frac{i^J 2^{2J-4}\Gamma(J-\frac{1}{2})}{\pi \Gamma(\frac{1}{2})}.
\end{gather}

\begin{figure}[H]
\center{\includegraphics[scale=0.8]{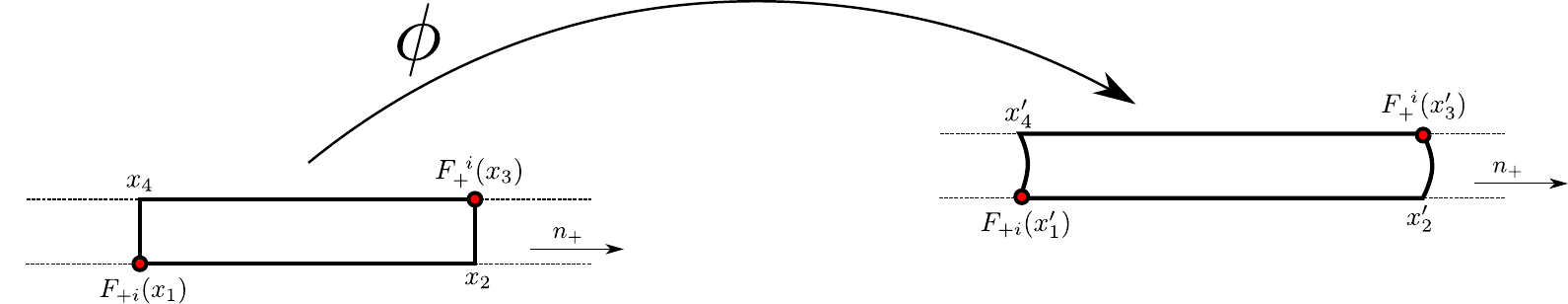} \\}
\caption{A generic conformal transformation \(\phi:\ x\rightarrow x'\) acting on infinitesimally narrow Wilson frame almost conserves its shape.    }
\label{ris:ConfPreobr}
\end{figure}

Finally, the superconformal operator with  conformal spin \(J\) reads as follows:
\begin{gather}
\breve{\mathcal{S}}^J(x_{1\bot})=\frac{i^J 2^{2J-4}\Gamma(J-\frac{1}{2})}{\pi \Gamma(\frac{1}{2})}\left(-(J-1)(J-2)\breve{S}_{sc}^J(x_{1\bot})
+i\frac{J-2}{2} \breve{S}_{f}^J(x_{1\bot})+\breve{S}_{gl}^J(x_{1\bot})\right).
\end{gather}
Now let us omit the common factor and redefine the operator as follows:
\begin{gather}
\breve{\mathcal{S}}^J(x_{1\bot})=-(J-1)(J-2)\breve{S}_{sc}^J(x_{1\bot})
+i\frac{J-2}{2} \breve{S}_{f}^J(x_{1\bot})+\breve{S}_{gl}^J(x_{1\bot}).\label{SPLL}
\end{gather}
In what follows, we will be interested in the BFKL limit when the Lorentz spin \(j\) goes to one: \(j=J-1=1+\omega\ \rightarrow 1\). In this limit, the operator takes the following form:
\begin{gather}
\breve{\mathcal{S}}^{2+\omega}(x_{1\bot})=-\omega \breve{S}_{sc}^{2+\omega}(x_{1\bot})+i\frac{\omega}{2}\breve{S}_{f}^{2+\omega}(x_{1\bot})+\breve{S}_{gl}^{2+\omega}(x_{1\bot}), \label{LR-1+omega}
\end{gather}
where we  kept in the coefficients only the leading order in \(\omega\).
Formally, such light ray operators are well defined objects. But in the
LLA BFKL the correlation functions  of two such objects, when  in each one  the local operators is placed at the same light ray, become ambiguous and should be regularized.     We will introduce the following regularization mentioned in the introduction: we replace
the light-ray operator (\ref{SPLL}) by a non-local  rectangular Wilson
loop with two opposite sides stretched along the light-cone direction and
two other sides (whose length tends to zero) being orthogonal to the light-cone. The fields are placed in two opposite corners of the Wilson frame. The configurations of Wilson frame and the positions of operators resulting from these operations are shown in Fig.\ref{ris:Ramka}.
  For example, for the gluons we get:
\begin{gather}
\breve{S}_{gl}^J(x_{1\bot})=\lim\limits_{(x_{1\bot}-x_{3\bot})^2\rightarrow 0} S_{gl}^J(x_{1\bot},x_{3\bot})= \notag\\
=\lim\limits_{(x_{1\bot}-x_{3\bot})^2\rightarrow 0}\int\limits_{-\infty}^{\infty} d x_{1-}\int\limits_{x_{1-}}^{\infty} dx_{3-} (x_{3-}-x_{1-})^{-J+2}\ \tr F^{\ \ i}_{+}(x_{1})[x_1,x_3]_{{}_\Box}F_{+i}(x_{3}),\label{RegulInOrthogSpace}\\
x_1=(x_{1-},0,x_{1\bot}),\ \ x_3=(x_{3-},0,x_{3\bot})\notag
\end{gather}
and the full regularized operator reds as follows:
\begin{gather}
\mathcal{S}^J(x_{1\bot},x_{3\bot})=-(J-1)(J-2)S_{sc}^J(x_{1\bot},x_{3\bot})
+i\frac{J-2}{2} S_{f}^J(x_{1\bot},x_{3\bot})+S_{gl}^J(x_{1\bot},x_{3\bot}),
\end{gather}
where \([x_1,x_3]_{{}_\Box}\) - Wilson frame with the local fields placed at the corners  \(x_1\), \(x_3\) on a diagonal of the frame and the short sides \(x_{23}\), \(x_{41}\) directed into the orthogonal space. The operation "\(\lim\)" is understood in the following  sense: at first we should carry out all calculation with the fixed length of short sides \(|x_{1\bot}-x_{3\bot}|^2\neq0\) , and only after that we take the limit. In this sense we can treat our infinitesimally narrow Wilson frame as a conformal object. Namely if we carry out any conformal transformation  this infinitesimally narrow Wilson frame almost conserves its shape \ref{ris:ConfPreobr}.

{\bf }

\section{ OPE over colour dipoles  for nonlocal operators $\mathcal{S}_{nloc}^{J}$}
Let us introduce two nonlocal super-primary operators defined above: the first one, \(\mathcal{S}_+^{2+\omega_1}\), is placed  along \(n_+\) and the second, \(\mathcal{S}_{-}^{2+\omega_2}\),  along \(n_-\). Our goal is to calculate their correlation function in the BFKL limit \(\omega_1,\omega_2 \rightarrow 0\), \(\frac{g^2}{\omega_1}, \ \frac{g^2}{\omega_2}\to \text{fixed}\). The main contribution in this case comes from  large  distances \(L_+\)(\(L_-\))   along \(n_+\)(\(n_-\)). The integral over \(L_+\)(\(L_-\))
entering into the definition of the regularized light-ray operator leads to the Regge pole \(\frac{1}{\o_{1,2}}\). This pole is analogous to the large \(\log\frac{s}{M^2}\) in the Regge approximation for high energy scattering amplitudes. \footnote{In high energy scattering \(M\) is a reference scale  such that \(m^2\ll M^2\ll s\), where \(m\) is a characteristic hadron mass. } Summing all contributions in \(\frac{g^2}{\o_{1,2}}\) in our setup corresponds to summing the powers \(g^2\log\frac{s}{M^2}\) in the LLA  in high energy scattering - the key feature   of the BFKL approximation.

In this case we can apply the OPE over color dipoles, which was elaborated for the scattering amplitudes in \cite{Balitsky:1995ub}.  Let us remind the logic of this approach within the scattering theory and then relate it to our calculation.

\begin{figure}[H]
\center{\includegraphics[scale=1.0]{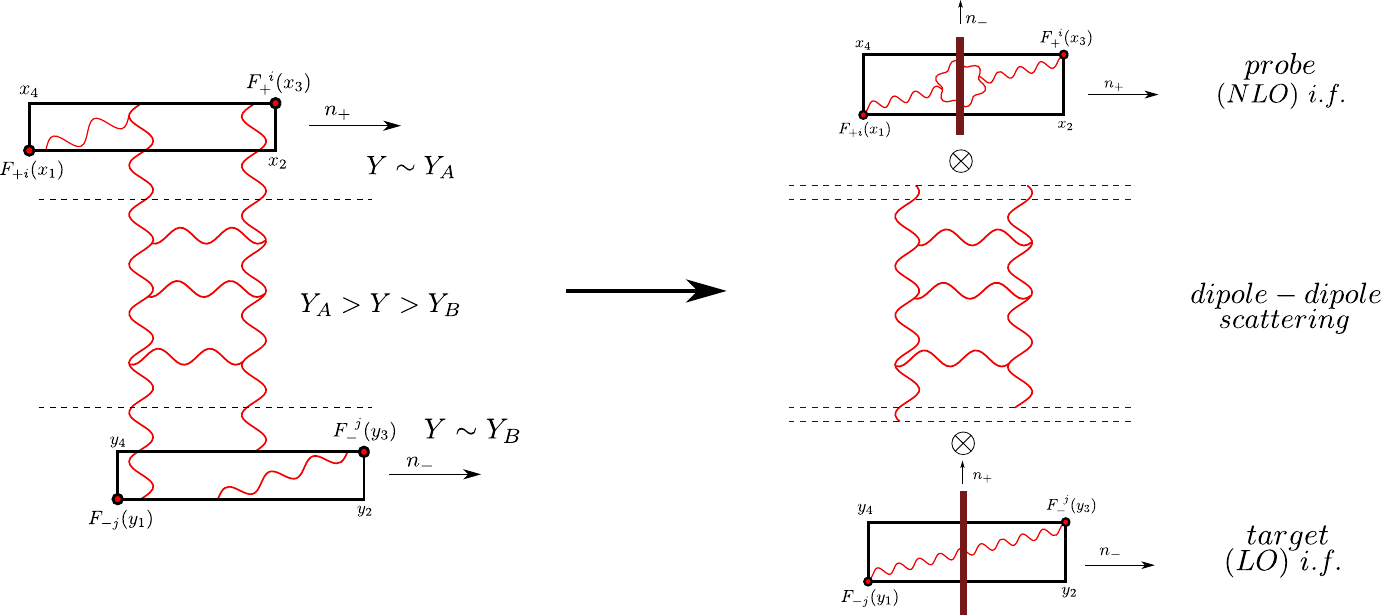}}
\caption{Colour dipole decomposition for the correlator of two frames. Due to the separation of scales w.r.t. the rapidity \(Y\) in BFKL approximation, the correlator factorizes into the ``probe impact factors", the dipole-dipole interaction and the ``target impact factor".   The analogue of rapidity \(Y\)   in the current paper is \(\log\s\) -- the logarithm of cutoff   for the momenta of  the gauge fields in the light cone direction.  As an example, the probe is represented by an NLO graph where as the target -- by an LO graph. }
\label{ris:Factorisation}
\end{figure}

The high-energy behavior of the amplitudes can be studied in the framework of the rapidity evolution of Wilson-line operators forming color dipoles.
The main idea is the factorization in rapidity: we separate a typical functional integral describing scattering of two particles into (i) the integral over the
gluon (and gluino) fields with rapidity close to the rapidity of the "probe" $Y_A$ , (ii) the integral over the gluons with rapidity close to the rapidity
of the target \(Y_B\), and (iii) the integral over the intermediate region of rapidities \(Y_A>Y>Y_B\)
 , see Fig. \ref{ris:Factorisation}. The result of the first integration is a certain coefficient function (impact factor) times color dipole (ordered in the
 direction of the probe velocity) with rapidities up to $Y_A$. Similarly, the result of the second integration is again the impact factor
 times the color dipole ordered in the direction of targetÕ velocity with rapidities greater than $Y_B$. The result of the last integration is the correlation function
 of two dipoles which can be calculated using the evolution equation for color dipoles,  known in the leading and next-to-leading order \cite{Balitsky:1995ub}.

To factorize in rapidity,  it is convenient to use the background field formalism: we integrate over gluons with $Y>Y_A$ and leave the gluons with $Y<Y_A$ as a background field, to
be integrated over later.  Since the rapidities of  background
gluons are very different from the rapidities of gluons in our Feynman diagrams, the background field is seen by the probe in the form of a shock wave (pancake) due to the Lorentz contraction.
To derive the expression of a quark or gluon propagator in this shock-wave background we represent the propagator as a path integral over various trajectories,
each of them weighed with the gauge factor Pexp$(ig\int\! dx_\mu A^\mu)$ ordered along the propagation path. Now, since the shock wave is very thin, quarks or gluons emitted by the probe do not
have time to deviate in transverse direction so their trajectory inside the shock wave can be approximated by a segment of the straight line. Moreover, since there is no external field
outside the shock wave, the integral over the segment of straight line can be formally extended to \(\pm\infty\) limits yielding the Wilson-line
gauge factor
\begin{gather}
U^{\sigma_+}_{x_\bot}=P\exp[i g_{{}_{YM}}
\int\limits_{-\infty}^\infty dx_+A^{\sigma_+}_-(x)],
\label{fund line factor}\end{gather}
where we have used the gauge field with a cutoff  \(\s=e^Y\) w.r.t. the longitudinal momenta \(k_+\)
\begin{gather}
A_{\mu}^{\sigma_+}(x)=\int d^4k \theta(\sigma_+-|k_+|)e^{ikx}A_\mu(k).
\label{A cutoff}\end{gather}
Now let us adopt this scheme of calculation to our correlator. In our case, the Wilson frames play the role of the probe and the target, respectively. The gluons emitted and absorbed within each frame contribute to their   "impact factors". The correlator factorizes into these two impact factors and the BFKL evolution  of colour dipoles  appearing in the OPE of the frames.      The BFKL evolution corresponds to the evolution of the cutoff from some minimal\footnote{The initial point $\sigma_0$ is an analog of the low normalization point $\mu^2\sim Q_0^2~\sim 1$GeV
for usual DGLAP evolution. It should be chosen in such way that $\sigma_0\gg M$ but $g^2\ln{\sigma_0\over M}\ll 1$ where M
is of order of the mass of colliding particles (in our case of M is of order of inverse transverse separations of Wilson frames).} \(\s_{0+}\) to the final value  \(\s_+\) for the  frame oriented in the \(n_-\) direction. Similarly, for the second dipole, the evolution goes from  \(\s_{0-}\) to  \(\s_{-}\). As we will see later, the ratio \(\frac{\s_{+}\s_{-}}{\s_{0+}\s_{0-}}\) will be identified with a certain anharmonic ratio of characteristic sizes of the configuration.  The rest of the calculation is very similar to the case of  scattering in Regge kinematics and is based on  computations of graphs in the pancake background.

The propagators of gluons, scalars and fermions get modified by the presence of this pancake background. Denoting the corresponding  average as  \(\langle\dots\rangle\) we represent these propagators  as follows\footnote{pancake is placed along \(n_+\) direction}:

\begin{gather}
\langle A_\mu^a(x)A_\nu^b(y)\rangle=\frac{1}{4\pi^3}\int d^{2}z_{\bot}U^{\sigma_+ab}_z[x_+g^\bot_{\mu\xi}-n_{-\mu}(x-z)_\xi^\bot][y_+\delta_\nu^{\bot\xi}-n_{-\nu}(y-z)^\xi_\bot]\cdot\notag\\
\cdot\frac{x_+|y_+|}{[-2(x-y)_- x_+ |y_+|+x_+(y-z)_\bot^2+|y_+|(x-z)_\bot^2+i\epsilon]},
\label{GlProp}
\end{gather}
\begin{gather}
\langle \hat{\Phi}^a_I(x)\hat{\Phi}^b_J(y)\rangle =
\frac{\delta_{IJ}}{4\pi^3}\int d^2z_\bot \frac{x_+|y_+| U^{ab}_z}{
\left[-2(x-y)_- x_+ |y_+|+(x-z)_\bot^2|y_+|+(y-z)_\bot^2x_++i\epsilon\right]^2},
\end{gather}

\begin{gather}
\langle \lambda^{aI}_\alpha(x)\,\bar\lambda^{bJ}_{\dot\alpha}(y)\rangle={i\over 2\pi^3}\int d^2z_\bot U^{ab}_z\left[x_+\bar{n}_-+(\bar{x}-\bar{z})\right]n_+\left[|y_+|\bar{n}_--(\bar{y}-\bar{z})\right]\cdot\notag\\
\cdot{x_+|y|_+\over [-2(x-y)_- x_+ |y_+|+(x-z)_\bot^2|y_+|+(y-z)_\bot^2 x_+
+i\epsilon]^3},\label{gluinoprop}
\end{gather}
where $\bar{n}_{\alpha\dot{\alpha}}\equiv n_\mu\bar{\sigma}^\mu_{\alpha\dot{\alpha}}$ and
$n^{\dot{\alpha}\alpha}\equiv n^\mu{\sigma}_\mu^{\dot{\alpha}\alpha}$, and \(U^{ab}_{z}=2\text{tr}(t^aU_{z}t^bU^\dagger_{z})\).
\par\medskip

The effective propagator for any field \(\chi\) has a form of decomposition over Wilson lines
\(\langle \chi^a(x)\chi^b(y)\rangle=\int d^2z_\bot U^{\sigma_+ ab}_{z_\bot}f(z,x,y)\), where \(f(z,x,y)\) is a function which depends only on  the coordinates and doesn't carry the colour indices. Then any conformal nonlocal operator \(S_{sc}^J\), \(S_{f}^J\), \(S_{gl}^J\) can be decomposed over the colour dipoles:
\begin{gather}
\text{tr(}\chi(x)[x,y]_{{}_\Box}\chi(y))\rightarrow \int d^2z_{\bot} f(z,x,y)U^{\sigma_+ ab}_{z_\bot}\text{tr}(t^a U^{\sigma_+}_{x\bot}t^bU^{\sigma_+}_{y\bot})\xrightarrow[N\rightarrow\infty]{}\notag\\
\xrightarrow[N\rightarrow\infty]{} \int d^2z_\bot f(z,x,y)\frac{N^2}{2}(1-\frac{1}{N}\text{tr}(1-U^{\sigma_+}_{x\bot}U^{\sigma_+\dag}_{z\bot})-\frac{1}{N}\text{tr}(1-U^{\sigma_+}_{z\bot}U^{\sigma_+\dag}_{y\bot})+O(g^2)), \label{decomposition}
\end{gather}
where we have used the following sequence of equalities:
\begin{gather}
U^{\sigma_+ ab}_{z}\text{tr}(t^a U^{\sigma_+}_{x}t^b U^{\sigma_+}_y)=2\text{tr}(t^a U^{\sigma_+}_z t^b U^{\sigma_+\dag}_z)\, \text{tr}( t^a U^{\sigma_+}_{x}t^b U^{\sigma_+}_y)=\text{tr}(t^a U^{\sigma_+}_xU^{\sigma_+\dag}_z t^a U^{\sigma_+}_z U_y^{\sigma_+\dag})=\notag\\
=\frac{1}{2}\text{tr}(U^{\sigma_+}_x U_z^{{\sigma_+}\dag})\,\text{tr}(U^{\sigma_+}_z U_y^{\sigma_+\dag})-\frac{1}{2N}\text{tr}(U^{\sigma_+}_x U_y^{\sigma_+\dag})\xrightarrow[N\rightarrow\infty]{}\notag\\
\xrightarrow[N\rightarrow\infty]{}\frac{N^2}{2}\left[1-\Ub^{\sigma_+}(x,z)-\Ub^{\sigma_+}(z,y)+\Ub^{\sigma_+}(x,z)\Ub^{\sigma_+}(z,y)\right]=\notag\\
=\frac{N^2}{2}\left[1-\Ub^{\sigma_+}(x,z)-\Ub^{\sigma_+}(z,y)\right]\left(1+O(g^2,\frac{1}{N^2})\right),
\label{color decomposition}\end{gather}
where we have introduced the colour dipole operator in fundamental representation:
\begin{gather}
\Ub^{\sigma_+}(x_{1\bot},z_{\bot})=1-\frac{1}{N}\text{tr}(U^{\sigma_+}_{x_{1\bot}}U^{\sigma_+\dag}_{z_{\bot}})\label{DipInFundRep}
\end{gather}
with  \(U^{\sigma_+}_{x_\bot}\)  defined in \eqref{fund line factor}.

The first two lines of  \eqref{color decomposition}  hold at any \(N\). In the  last line  we dropped the   term  non-linear in colour dipoles. This is valid since in the BFKL approximation we take into account only linear evolution of Wilson-line operators corresponding to the processes containing two reggeized gluons in t-channel. \footnote{The nonlinear
 terms are relevant for the high-energy in dense QCD regime like $pA$ scattering on LHC}

Let us now  proceed with calculation of the  gluonic part. Calculation for scalars and fermions can be done in the same way and it turns out that the main contribution for the ``impact factor" in LO comes just from gluons.  Naively, it can be explained from the fact that in the limit \(\o=J-2\to 0\) the scalar and fermionic terms enter with subleading coefficients into the eq.\eqref{LR-1+omega}. The explicit computation of correlators confirms it.

Using the propagator (\ref{GlProp}) and the decomposition (\ref{decomposition}) the OPE for the gluon operator  stretched in \(n_+\) direction reads as follows\footnote{The representation for the impact factor as an integral of 4-point correlator of the external currents and the gluonic current was constructed in \cite{Cornalba:2008qf,Cornalba:2009ax}}

\begin{gather}
\text{tr\,} F_{+i}(x_1)[x_1,x_3]F^{\ \ i}_+(x_3) \rightarrow  \notag\\
\frac{N^2}{2\pi^3}\int d^2z \left( \frac{2}{(x_{3-}(x_1-z)_\bot^2-x_{1-}(x_3-z)_\bot^2)^2}+\frac{x_{1-}x_{3-}(9(x_1-z)_\bot^2(x_3-z)_\bot^2+6(x_1-z,x_3-z)^2)}{(x_{3-}(x_1-z)_\bot^2-x_{1-}(x_3-z)_\bot^2)^4} \cdot \right. \notag \\
\left. (1-\Ub^{\sigma_-}(x_{1\bot},z_{\bot})-\Ub^{\sigma_-}(z_{\bot},x_{3\bot})) \right),\label{FFpropagator}
\end{gather}
where we use in \(\Ub^{\s_-}\) the gauge field with a cut-off \(\s_-\) for the light-cone momenta.
Now we can collect the full expression for the correlator of   operators \(S_{gl}^{2+\omega_1}\) and \(S_{gl}^{2+\omega_2}\)  stretched along \(n_+\) and \(n_-\) directions using \eqref{RegulInOrthogSpace} and \eqref{FFpropagator}:
\begin{gather}
\langle S_{gl}^{2+\omega_1}(x_{1\bot},x_{3\bot})S_{gl}^{2+\omega_2}(y_{1\bot},y_{3\bot})\rangle=\notag\\
=(\frac{N^2}{2\pi^3})^2\int\limits_{-\infty}^{\infty}dx_{1-} \int\limits_{x_{1-}}^{\infty}dx_{3-}(x_{3-}-x_{1-})^{-\omega_1} \int\limits_{-\infty}^{\infty}dy_{1+} \int\limits_{y_{1+}}^{\infty}dy_{3+}(y_{3+}-y_{1+})^{-\omega_2} \cdot \notag\\
\cdot \int d^2z \left( \frac{2}{((x_{3-}(x_1-z)_\bot)^2-x_{1-}(x_3-z)_\bot^2)^2}+\frac{x_{1-}x_{3-}(9(x_1-z)_\bot^2(x_3-z)_\bot^2+6(x_1-z,x_3-z)^2)}{(x_{3-}(x_1-z)_\bot^2-x_{1-}(x_3-z)_\bot^2)^4}\right)\notag\\
\cdot \int d^2w \left( \frac{2}{((y_{3+}(y_1-w)_\bot)^2-y_{1+}(y_3-w)_\bot^2)^2}+\frac{y_{1+}y_{3+}(9(y_1-w)_\bot^2(y_3-w)_\bot^2+6(y_1-w,y_3-w)^2)}{(y_{3+}(y_1-w)_\bot^2-y_{1+}(y_3-w)_\bot^2)^4}\right)\notag\\
(\langle \Ub^{\sigma_-}(x_{1\bot},z_{\bot})\Vb^{\sigma_+}(y_{1\bot},w_{\bot}) \rangle + \langle \Ub^{\sigma_-}(x_{1\bot},z_{\bot})\Vb^{\sigma_+}(w_{\bot},y_{3\bot})\rangle+\notag\\
\langle \Ub^{\sigma_-}(z_{\bot},x_{3\bot})\Vb^{\sigma_+}(y_{1\bot},w_{\bot}) \rangle + \langle \Ub^{\sigma_-}(z_{\bot},x_{3\bot})\Vb^{\sigma_+}(w_{\bot},y_{3\bot})\rangle),\label{gluon-gluon}
\end{gather}
where \(\Vb^{\sigma_+}(y_{1\bot},w_{\bot})\) is the operator similar to (\ref{DipInFundRep}) but for the second frame operator stretched along \(n_-\).

All terms in the last brackets are similar and give the same contribution. We proceed with the first one. As was demonstrated in \cite{Balitsky:1995ub}, the problem of calculation of correlator for two dipoles splits into two  parts: we      compute the correlator for relatively small values of the cutoff \(\tilde{\sigma}\) such that \(g^2\log\frac{\tilde{\sigma}}{\sigma_{_0}}\ll 1\), where \(\sigma_0\ < \tilde{\sigma}\ll \sigma\) for each dipole, when the lowest order of perturbation theory dominates in the LLA BFKL approximation, and then we evolve the result w.r.t. \(\tilde{\sigma}\) to its final value \(\sigma\).  Let us elaborate it in detail.

\subsection{BFKL evolution}
As was demonstrated in \cite{Balitsky:2009xg} evolution w.r.t. the cutoff can be written in the form of BFKL equation\footnote{we dropped index \(\bot\) for sake of brevity. Also we will omit index \(\bot\) in all formulas where it doesn't lead to confusion.}:
\begin{gather}
\sigma \frac{d}{d\sigma}\Ub^\sigma(z_1,z_2)=\mathcal{K}_{{\rm BFKL}}\ast \Ub^\sigma(z_1,z_2), \label{BFKL}
\end{gather}
where \(\mathcal{K}_{\rm BFKL}\) is the integral operator having the following form in LO BFKL approximation:
\begin{equation}
\mathcal{K}_{_{\rm LO\, BFKL}}\ast\Ub(z_1,z_2)=\frac{2g^2}{\pi} \int d^2z_3\frac{z_{12}^2}{z_{13}^2z_{23}^2}\left[\Ub(z_1,z_3)+
\mathcal{\Ub}(z_3,z_2)-\mathcal{\Ub}(z_1,z_2)\right].
\end{equation}
In principal, we will use in what follows the NLO generalization of this kernel, or rather of its eigenvalues, to fix the  NLO  scaling of the correlator. To fix the right NLO normalization of the correlator, we should also correct the operators \(\Ub,\Vb\), but we will restrict ourselves to the LO in the normalization.

The BFKL kernel has \(G=SL(2,C)\) symmetry and its  eigenfunctions and the spectrum should be classified w.r.t. the irreps of this group. The \(SL(2,C)\) group has three sets of unitary irreps. The color dipole operator can be expanded w.r.t. only one  of them, the   principal series, characterised by  conformal weights \(h=\frac{1+n}{2}+i\nu\), \(\bar{h}=\frac{1-n}{2}+i\nu\), where \(\nu\in\mathbb{R}\), \(n\in \mathbb{Z}\) and a two-dimensional coordinate \(z_0\). Explicitly, the eigenfunction reads as follows\footnote{where we pass to the complex coordinates for  2-dimensional space: \(z=(x,y)\rightarrow x+iy\). Complex conjugation of \(z\) is denoted by \(\bar{z}\) }:
\begin{gather}
E_{h,\bar{h}}(z_{10},z_{20})=\left[\frac{z_{12}}{z_{10}z_{20}}\right]^{h}\left[\frac{\bar{z}_{12}}{\bar{z}_{10}\bar{z}_{20}}\right]^{\bar{h}}.
\end{gather}
Let us introduce the projection of dipole on the \(E\)-function:
\begin{gather}
\mathcal{U}_{\nu,n}(z_0)=\frac{1}{\pi^2}\int \frac{d^2z_1d^2z_2}{z_{12}^4}E^\ast_{\nu,n}(z_{10},z_{20})\Ub(z_1,z_2)\label{ProjectedDipole}
\end{gather}
and the  inverse transformation is
\begin{gather}
\Ub(z_1,z_2)=\sum\limits_{n=-\infty}^{\infty}\int d\nu\int d^2z_0\frac{\nu^2+n^2/4}{\pi^2}E_{\nu,n}(z_{10},z_{20})\mathcal{U}_{\nu,n}(z_0). \label{inverse formula}
\end{gather}
The solutions of BFKL equation  (\ref{BFKL})  in terms of this projection   can be explicitly written in the following form:
\begin{gather}
\mathcal{U}^\sigma_{\nu,n}(z_0)=\left(\frac{\sigma}{\sigma_0}\right)^{\aleph(\nu,n)}\mathcal{U}^{\sigma_0}_{\nu,n}(z_0), \label{BFKLevolProj}
\end{gather}
where \(\aleph(n,\nu)\) are the  eigenvalues of \(\mathcal{K}_{BFKL}\). Let us give their expression already in the NLO approximation \cite{Kotikov:2002ab}
\begin{gather}
\aleph(\nu,n=0)=4g^2(\chi(\nu)+g^2\delta(\nu)),\notag\\
\chi(\nu)=2\Psi(1)-\Psi(\frac{1}{2}+i\nu)-\Psi(\frac{1}{2}-i\nu),\notag\\
\delta(\nu)=\chi''(\nu)+6\zeta(3)-2\zeta(2)\chi(\nu)-2\Phi(\frac{1}{2}+i\nu)-2\Phi(\frac{1}{2}-i\nu), \label{Aleph}
\end{gather}
where \(\Psi(x)=\frac{\Gamma'(x)}{\Gamma(x)}\) and function \(\Phi(x)\) has the following representation:
\begin{gather}
\Phi(x)=\frac{1}{2}\sum\limits_{k=0}^\infty \frac{\Psi'(\frac{k+2}{2})-\Psi'(\frac{k+1}{2})}{k+x}.
\end{gather}
The transformation (\ref{ProjectedDipole}) can be expressed graphically as in Fig.\ref{ris:DipProj}:
\begin{figure}[H]
\center{\includegraphics[scale=1.1]{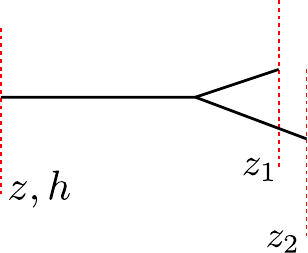} \\}
\caption{Graphical representation of  the  projection ~  (\ref{ProjectedDipole}) of colour dipole on the \(E\)-eigenfunction. Dotted lines represent the Wilson lines and all the coordinates correspond to the transverse 2-dimensional space.}
\label{ris:DipProj}
\end{figure}
\subsection{Correlator of dipoles with a small cutoff}
\begin{figure}[H]
\center{\includegraphics[scale=0.8]{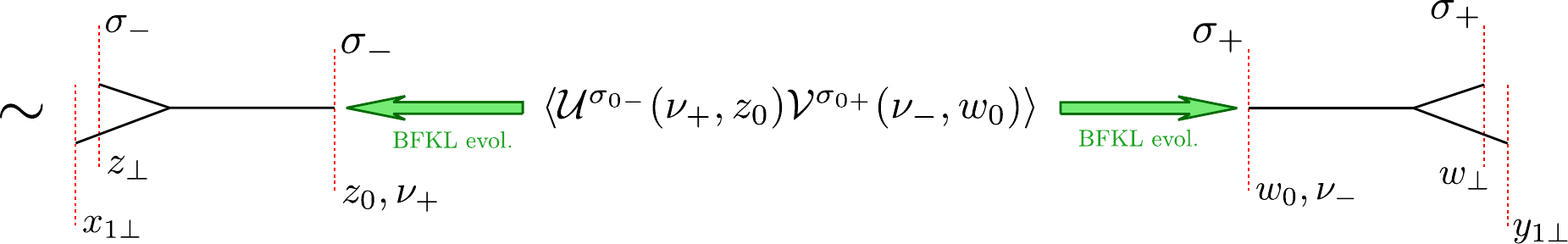} \\}
\caption{The logic of our calculation of the dipole-dipole correlation function: the projection of the colour dipoles onto the \(E\)-functions at each end-point, the BFKL evolution from relatively small cutoffs (green arrows) and, finally, the calculation of the dipole-dipole correlation function at small cutoffs, in the middle.  }
\label{ris:LogicOfCalcul}
\end{figure}

\begin{figure}[H]
\center{\includegraphics[scale=0.77]{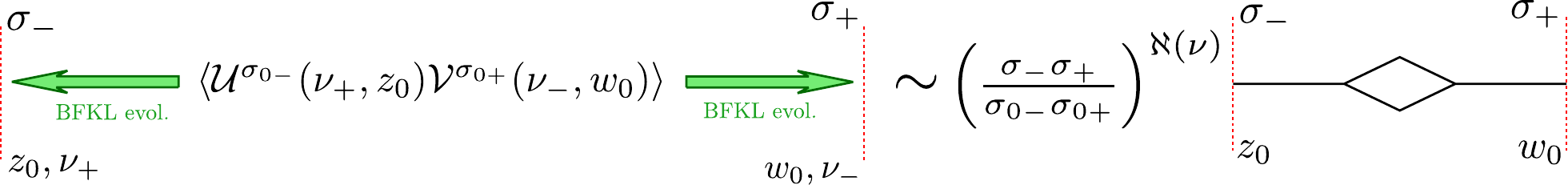} \\}
\caption{The scheme  of calculation of the dipole-dipole correlator for small cutoffs and the BFKL evolution (shown by green arrows). In the r.h.s. we use the orthogonality condition for the \(E\)-functions. }
\label{ris:SmallCUTOFF}
\end{figure}

Now let us introduce projections of \(\Ub^{\sigma_-}(x_{1\bot},z_{\bot})\) and \(\Vb^{\sigma_+}(y_{1\bot},w_{\bot})\) to the eigenfunctions:
\begin{gather}
\mathcal{U}_{\nu_+,n_+}^{\sigma_-}(z_0)=\frac{1}{\pi^2}\int \frac{d^2x_{1\bot}d^2z_\bot}{(|x_{1\bot}-z_\bot|^2)^2}E^\ast_{\nu_+,n_+}(x_{1\bot}-z_0,z_\bot-z_0)\Ub^{\sigma_-}(x_{1\bot},z_{\bot}),\\
\mathcal{V}_{\nu_-,n_-}^{\sigma_+}(w_0)=\frac{1}{\pi^2}\int \frac{d^2y_{1\bot}d^2w_\bot}{(|y_{1\bot}-w_\bot|^2)^2}E^\ast_{\nu_-,n_-}(y_{1\bot}-w_0,w_\bot-w_0)\Vb^{\sigma_+}(y_{1\bot},w_{\bot}).
\end{gather}
It is important to stress that the contribution corresponding to the lowest twist-2 comes from the projections with \(n=0\). Using BFKL evolution (\ref{BFKLevolProj}) we can reduce the correlator with arbitrary cut-off to the case of  small\footnote{But large enough to use LLA and only two-reggion contribution to \(\aleph(n,\nu)\)} cutoffs \(\s_{\pm0}\):
\begin{gather}
\langle \mathcal{U}_{\nu_+}^{\sigma_-}(z_0),\mathcal{V}_{\nu_-}^{\sigma_+}(w_0) \rangle=\left(\frac{\sigma_{-}}{\sigma_{-0}}\right)^{\aleph(\nu_+)}\left(\frac{\sigma_{+}}{\sigma_{+0}}\right)^{\aleph(\nu_-)}\langle \mathcal{U}_{\nu_+}^{\sigma_{0-}}(z_0),\mathcal{V}_{\nu_-}^{\sigma_{0+}}(w_0) \rangle. \label{CorForProj}
\end{gather}
Graphically the logic of our calculation can be represented as in Fig.\ref{ris:LogicOfCalcul}

\begin{gather}
\langle \Ub^{\sigma_-}(x_{1\bot},z_{\bot})\Vb^{\sigma_+}(y_{1\bot},w_{\bot}) \rangle \sim \notag
\end{gather}
The correlation function between two dipoles with relatively small cutoffs \(\s_\pm\gtrsim\s_{\pm0}\) can be calculated perturbatively. In one loop it reads as follows \cite{Balitsky:2009xg}
\begin{eqnarray}
&&\langle \mathcal{U}_{\nu_+}^{\sigma_{-}}(z_0),\mathcal{V}_{\nu_-}^{\sigma_{+}}(w_0) \rangle =\frac{-4\pi^4g^4}{N^2\nu_-^2(\nu_-^2+\frac{1}{4})^2}\times\notag\\
&&\left(\delta(z_0-w_0)\delta(\nu_++\nu_-)+\frac{2^{1-4i\nu_-}\delta(\nu_+-\nu_-)}{\pi|z_0-w_0|^{2-4i\nu_-}}\frac{\Gamma(\frac{1}{2}+i\nu_-)\Gamma(1-i\nu_-)}{\Gamma(i\nu_-)\Gamma(\frac{1}{2}-i\nu_-)}  \right)(1+O\left(g^2\log\left(\frac{\sigma_-\sigma_+}{\sigma_{0-}\sigma_{0+}}\right)\right).\notag\\
\label{TreeLevelCor}
\end{eqnarray}
Graphically, the last calculation, together with the BFKL evolution,  looks as depicted in the Fig.\ref{ris:SmallCUTOFF}
\section{Calculation of correlation function}
Now using the inversion formula (\ref{inverse formula}) and the eq.(\ref{CorForProj}) we obtain the correlator of two color dipoles with the original finite cut-offs \(\s_{\pm}\):
\begin{gather}
\langle \Ub^{\sigma_-}(x_{1\bot},z_{\bot}) \Vb^{\sigma_+}(y_{1\bot},w_{\bot})\rangle = -\frac{4\pi^2g^4}{N^2}\int d\nu_+ \int d^2 z_0 \frac{\nu_+^2}{\pi^2}\left(\frac{(x_1-z)_{\bot}^2}{(x_1-z_0)_{\bot}^2(z-z_0)_{\bot}^2} \right)^{\frac{1}{2}+i\nu_+}\cdot\notag\\
\cdot\int d\nu_- \int d^2 w_0 \frac{\nu_-^2}{\pi^2}\left(\frac{(y_1-w)_{\bot}^2}{(y_1-w_0)_{\bot}^2(w-w_0)_{\bot}^2} \right)^{\frac{1}{2}+i\nu_-}
\left(\frac{\sigma_{-}}{\sigma_{0-}}\right)^{\aleph(\nu_+)}\left(\frac{\sigma_{+}}{\sigma_{0+}}\right)^{\aleph(\nu_-)}\frac{\pi^2}{\nu_-^2(\nu_-^2+\frac{1}{4})^2}\cdot\notag\\
\cdot\left(\delta(z_0-w_0)\delta(\nu_++\nu_-)+\frac{2^{1-4i\nu_-}\delta(\nu_+-\nu_-)}{\pi|z_0-w_0|^{2-4i\nu_-}}\frac{\Gamma(\frac{1}{2}+i\nu_-)\Gamma(1-i\nu_-)}{\Gamma(i\nu_-)\Gamma(\frac{1}{2}-i\nu_-)}\right).
\end{gather}
Integrating over \(\nu_-\) and\footnote{using the "star-triangle" relation, see e.g. the Appendix A of \cite{Derkachov:2001yn}} over \(w_0\) we get:
\begin{gather}
\langle \Ub^{\sigma_-}(x_{1\bot},z_{\bot}) \Vb^{\sigma_+}(y_{1\bot},w_{\bot})\rangle = \notag\\
=-\frac{8g^4}{N^2}\int  \int \frac{  d\nu\,\nu^2\,\,d^2 z_0}{(\frac{1}{4}+\nu^2)^2}\left(\frac{(x_1-z)_{\bot}^2}{(x_1-z_0)_{\bot}^2(z-z_0)_{\bot}^2} \right)^{\frac{1}{2}+i\nu}\left(\frac{(y_1-w)_{\bot}^2}{(y_1-z_0)_{\bot}^2(w-z_0)_{\bot}^2} \right)^{\frac{1}{2}-i\nu}
\left(\frac{\sigma_+\sigma_-}{\sigma_{+0}\sigma_{-0}}\right)^{\aleph(\nu)}.
\label{corrUV}\end{gather}
Now we come to the  subtlest point in our calculations:  we should fix the physical value of   the ratio  of our cutoffs \(\frac{\sigma_+\sigma_-}{\sigma_{-0}\sigma_{+0}}\).
In general, they are some functions of our configuration of the frames and should depend on conformal ratios of 8 points characterizing the shapes and positions of  two frames. But we expect  that in the limit of very narrow frames which we will need,  the cutoffs depend  only on the  distances between the positions of local fields \(x_1\), \(x_3\), \(y_1\), \(y_3\). Indeed, if we make a conformal transformation with generic parameters, which are not related to the  shape of the frames, the frames remain rectangular (up to an insignificant, in our limit, deformation of their short sides, as shown in Fig.\ref{ris:2Ramki}) and and are still characterized  by 4 points \(x_1'\), \(x_3'\), \(y_1'\), \(y_3'\) -  new  position of the local fields inserted into the frames. That means that the cutoffs can depend only on two conformally invariant ratios: \(r_1=\frac{(x_1-y_3)^2(x_{3}-y_1)^2}{x_{13}^2y_{13}^2}\) and  \(r_2=\frac{(x_1-y_1)^2(x_{3}-y_3)^2}{x_{13}^2y_{13}^2}\) and we have to calculate this dependence explicitly.  A natural assumption would be then to put \(\frac{\sigma_+\sigma_-}{\sigma_{-0}\sigma_{+0}}\simeq r_2\) or \(\frac{\sigma_+\sigma_-}{\sigma_{-0}\sigma_{+0}}\simeq r_1\). Both choices give the same result in the limit when the distance between the frames is much less then their lengths. It is demonstrated in the appendix \ref{CutOFFfromNLO}  that precisely this cut-off  occurs in the NLO graphs in this limit. But before this limit the result appears to be a bit subtler.
Instead of doing explicit calculations, we will appeal to a similar calculation done in   in  \cite{Balitsky:2009yp}  for the 4-point correlator of local scalar fields  \(\langle\tr Z^2(x_1) \bar{Z}^2(x_3) Z^2(y_1) \bar{Z}^2(y_3)\rangle\). Its form   is fixed by conformal symmetry so that it depends on the same conformal ratios as our configuration. Comparing the result of BFKL computation of this correlator, which also uses the  OPE for the regularized colour dipoles,  with  the Regge limit of the same quantity found in the papers  \cite{Cornalba:2007fs,Costa:2012cb}, the following prescription was found for the cutoff dependence on \(x_1,x_3,y_1,y_3\):\footnote{By keeping each factor in the anharmonic ratio in a separate power \(\frac{\aleph(\nu)}{2}\) we choose the right analytic branch giving the right signature factor. It could be explicited by an \(i0\) prescription.}
\begin{gather}
\left(\frac{\sigma_+\sigma_-}{\sigma_{+0}\sigma_{-0}}\right)^{\aleph(\nu)} \rightarrow \notag\\
\rightarrow\frac{i}{\sin \pi \aleph(\nu)}
\left(\frac{((x_1-y_3)^2)^{\frac{\aleph(\nu)}{2}}((x_{3}-y_1)^2)^{\frac{\aleph(\nu)}{2}}}{(x_{13}^2)^{\frac{\aleph(\nu)}{2}}(y_{13}^2)^{\frac{\aleph(\nu)}{2}}}-
\frac{((x_1-y_1)^2)^{\frac{\aleph(\nu)}{2}}
((x_{3}-y_3)^2)^{\frac{\aleph(\nu)}{2}}}{(x_{13}^2)^{\frac{\aleph(\nu)}{2}}(y_{13}^2)^{\frac{\aleph(\nu)}{2}}}\right).
\label{cutoff}
\end{gather}
Since we are in a very similar kinematic situation we will assume here that our cutoff dependence on the coordinates of the frames is also governed by \eqref{cutoff}.

A motivation stemming from the Feynman graphs of NLO impact factor is given in Appendix \ref{CutOFFfromNLO}.

We denote  the coordinates of the vertices of a  frame as follows:
\begin{eqnarray}
x_1&=&(-uL_-,0,x_{1\bot}),\notag\\
x_3&=&(\bar{u}L_-,0,x_{3\bot}),\notag\\
y_1&=&(0,-v L_+,y_{1\bot}),\notag\\
y_3&=&(0,\bar{v}L_+,y_{3\bot}),
\end{eqnarray}
where \(\bar{u}=1-u,\ \bar{v}=1-v\). These parameters can be restricted to \(u,v \in (0,1)\) since each frame should intersect the shock wave in order to give a non-zero contribution. If there is no intersection, we can make a scale transformation which sends the frame to an infinite distance from the shock wave, thus suppressing their interaction.

   Using  (\ref{cutoff}) we can rewrite the correlator of two dipoles  \eqref{corrUV} in the explicit way:
\begin{gather}
\langle \Ub^{\sigma_-}(x_{1\bot},z_{\bot}) \Vb^{\sigma_+}(y_{1\bot},w_{\bot})\rangle=\notag\\
=-\frac{8g^4}{N^2}\int d\nu \int d^2 z_0 \frac{\nu^2}{(\frac{1}{4}+\nu^2)^2}\left(\frac{(x_1-z)_{\bot}^2}{(x_1-z_0)_{\bot}^2(z-z_0)_{\bot}^2} \right)^{\frac{1}{2}+i\nu}\left(\frac{(y_1-w)_{\bot}^2}{(y_1-z_0)_{\bot}^2(w-z_0)_{\bot}^2} \right)^{\frac{1}{2}-i\nu}\frac{i}{\sin\pi \aleph(\nu)}\cdot\notag\\
\cdot\left(\left(\frac{(2uL_-\bar{v}L_++\Delta^2_\bot)(2\bar{u}L_-vL_++\Delta^2_\bot)}{x^2_{13\bot}y^2_{13\bot}}\right)^{\frac{\aleph(\nu)}{2}}-\left(\frac{(-2uL_-vL_++\Delta^2_\bot)(-2\bar{u}L_-\bar{v}L_++\Delta^2_\bot)}{x^2_{13\bot}y^2_{13\bot}}\right)^{\frac{\aleph(\nu)}{2}}\right),
\end{gather}
where \(\D_\bot=(x-y)_\bot\) is the distance between the frames in the orthogonal direction.
Then we plug it into (\ref{gluon-gluon}) and thus obtain a closed expression for the correlator. Now we should  carry out the remaining integrations. Let us start with the  integrations over \(L_-\) and \(L_+\). We can factor out the L-dependence  in each of the two terms in (\ref{cutoff})    leading to the  following two  terms in (\ref{gluon-gluon}):
\begin{gather}
\int\limits_0^\infty dL_- L_-^{-2-\omega_1}\int\limits_{-L_-}^0dx_{1+}\int\limits_0^\infty dL_+ L_+^{-2-\omega_2}\int\limits_{-L_+}^0dy_{1-}\left((2uL_-\bar{v}L_++\Delta^2_\bot)(2\bar{u}L_-vL_++\Delta^2_\bot)\right)^{\frac{\aleph(\nu)}{2}}=\notag\\
=2\pi \delta(\omega_1-\omega_2)\int\limits_0^1\int\limits_0^1 du dv(4u\bar{u}v\bar{v})^{\frac{\aleph(\nu)}{2}}(\frac{\Delta^2}{2u\bar{v}})^{\frac{\aleph(\nu)}{2}}(\frac{\Delta^2}{2\bar{u}v})^{-\omega+\frac{\aleph(\nu)}{2}}\times\notag\\
\times B(-\omega,\omega-\aleph(\nu)){ }_2F_1(-\frac{\aleph(\nu)}{2},-\omega;-\aleph(\nu);1-\frac{u\bar{v}}{\bar{u}v})
\end{gather}
and
\begin{gather}
\int\limits_0^\infty dL_- L_-^{-2-\omega_1}\int\limits_{-L_-}^0dx^+\int\limits_0^\infty dL_+ L_+^{-2-\omega_2}\int\limits_{-L_+}^0dy^-\left((-2uL_-vL_++\Delta^2_\bot\right)^{\frac{\aleph(\nu)}{2}}\left(-2\bar{u}L_-\bar{v}L_++\Delta^2_\bot)\right)^{\frac{\aleph(\nu)}{2}}=\notag\\
=2\pi e^{i\pi\aleph(\nu)}(-1)^{\aleph(\nu)-\omega}\delta(\omega_1-\omega_2)\int\limits_0^1\int\limits_0^1 du dv(4u\bar{u}v\bar{v})^{\frac{\aleph(\nu)}{2}}(\frac{\Delta^2}{2uv})^{\frac{\aleph(\nu)}{2}}(\frac{\Delta^2}{2\bar{u}\bar{v}})^{-\omega+\frac{\aleph(\nu)}{2}}\times\notag\\
\times B(-\omega,\omega-\aleph(\nu)){ }_2F_1(-\frac{\aleph(\nu)}{2},-\omega;-\aleph(\nu);1-\frac{uv}{\bar{u}\bar{v}}).
\end{gather}
The best way to do these integrals is to change the integration variables to \(L_+L_-\) and \(\frac{L_+}{L_-}\). The integral over \(\frac{L_+}{L_-}\) renders \(\delta(\omega_1-\omega_2)\).    Here \(B(-\omega,\omega-\aleph(\nu))=\frac{\Gamma(-\omega)\Gamma(\omega-\aleph(\nu))}{\Gamma(-\aleph(\nu))}\), and thus it has a pole in \(\nu\), namely \(\frac{1}{\omega-\aleph(\nu)}\). We postpone the \(\nu\)-integration because we close the contour integration (in the upper or lower half-plane) depending on whether the modulo of the ratio
\begin{gather}
\left|\left(\frac{(x_1-z)^2}{(x_1-z_0)^2(z-z_0)^2} \right)\left(\frac{(y_1-w)^2}{(y_1-z_0)^2(w-z_0)^2} \right)^{-1}\right|
\end{gather}
is greater or smaller then one. Hence we first carry out the coordinate integrations. First let us perform the \(u\)- and \(v\)-integrations. We can factor out all functions  depending on \(u\) and do the \(u\)-integration:
\begin{gather}
\int \limits_0^1du (u(1-u))^{\frac{\aleph(\nu)}{2}}\frac{1}{u^{\frac{\aleph(\nu)}{2}}}\frac{1}{(1-u)^{\frac{\aleph(\nu)}{2}-\omega}}{ }_2F_1(-\frac{\aleph(\nu)}{2},-\omega;-\aleph(\nu);1-\frac{u\bar{v}}{\bar{u}v})\cdot\notag\\
\cdot \left( \frac{2}{(((1-u)(x_1-z)_\bot)^2+u(x_3-z)_\bot^2)^2}-\frac{u(1-u)(9(x_1-z)_\bot^2(x_3-z)_\bot^2+6(x_1-z,x_3-z)_{\bot}^2)}{((1-u)(x_1-z)_\bot^2+u(x_2-z)_\bot^2)^4}\right)=\notag\\
=\frac{1}{2}\left(\frac{1}{|x_3-z|_{\bot}^2|x_1-z|_{\bot}^2}-\frac{2[(x_3-z)_{\bot}\cdot(x_1-z)_{\bot}]^2}{(|x_3-z|_{\bot}^2|x_1-z|_{\bot}^2)^2} \right)(1+O(g^2,\omega)),
\end{gather}
where by \(\cdot\) we denote the scalar product of 2-dimensional vectors. The first line of the integrand is equal to \(1+O(g^2,\omega)\) and the terms  of the order \(O(g^2,\omega)\) contribute only to the NLO impact factor. We will drop such terms since we limit ourselves  to the LO to the impact factor only. Finally we obtain  our correlator of the regularized light ray operators in the form:
\begin{gather}
\langle S_{gl+}^{2+\omega_1}(x_{1\bot},x_{3\bot})S_{gl-}^{2+\omega_2}(y_{1\bot},y_{3\bot})\rangle=\notag\\
=-i\frac{4N^2g^4}{\pi^5}\delta(\omega_1-\omega_2)\int d \nu \frac{(\Delta_{\bot}^2)^{\aleph(\nu)-\omega}}{(x_{13\bot}^2y_{13\bot}^2)^{\frac{\aleph(\nu)}{2}}}B(-\omega,\omega-\aleph(\nu))\frac{1-e^{i\pi(2\aleph(\nu)-\omega)}}{\sin \pi \aleph(\nu)}\frac{\nu^2}{(\frac{1}{4}+\nu^2)^2}\cdot\notag\\
\cdot \int d^2z \left(\frac{1}{2|x_1-z|_{\bot}^2|x_3-z|_{\bot}^2}-\frac{(x_1-z,x_3-z)_{\bot}^2}{(|x_1-z|_{\bot}^2|x_3-z|_{\bot}^2)^2}\right)\cdot \notag\\\cdot\int d^2w \left(\frac{1}{2|y_1-w|_{\bot}^2|y_3-w|_{\bot}^2}-\frac{(y_1-w,y_3-w)_{\bot}^2}{(|y_1-w|_{\bot}^2|y_3-w|_{\bot}^2)^2}\right)\cdot\notag\\
\cdot\int d^2 z_0 \left(\frac{|x_1-z|_{\bot}^2}{|x_1-z_0|_{\bot}^2|z-z_0|_{\bot}^2} \right)^{\frac{1}{2}+i\nu}\left(\frac{|y_1-w|_{\bot}^2}{|y_1-z_0|_{\bot}^2|w-z_0|_{\bot}^2} \right)^{\frac{1}{2}-i\nu}.
\end{gather}
To be able to  calculate these integrals over \(z\) and \(w\) we derived, using the dimensional regularization and Feynman parameterization, the following formula:
\begin{gather}
2\int \frac{d^2 z}{\pi} \left( \frac{1}{(x-z)^2(y-z)^2}-\frac{2\langle x-z,y-z\rangle^2}{((x-z)^2(y-z)^2)^2}\right)
\frac{(x-z)^{2\b}}{z^{2\b}}
=-\frac{\b}{1-\b}\frac{2\langle x,y\rangle^2-x^2 y^2}{x^2(y^2)^{1+\b}((x-y)^2)^{1-\b}}\,.
\end{gather}
It leads to the following expression for our correlator:
\begin{gather}
\langle S_{gl+}^{2+\omega_1}(x_{1\bot},x_{3\bot})S_{gl-}^{2+\omega_2}(y_{1\bot},y_{3\bot})\rangle=\notag\\
=-i\frac{N^2g^4}{4\pi^3}\delta(\omega_1-\omega_2)\int d \nu (\Delta_{\bot}^2)^{\aleph(\nu)-\omega}B(-\omega,\omega-\aleph(\nu))\frac{1-e^{i\pi(2\aleph(\nu)-\omega)}}{\sin \pi \aleph(\nu)}\frac{\nu^2}{ (\frac{1}{4}+\nu^2)^2}\cdot\notag\\
\frac{1}{(|x_{13}|_{\bot}^2|y_{13}|_{\bot}^2)^{\frac{1}{2}+\frac{\aleph(\nu)}{2}}}\int d^2z_0\frac{(|x_{13}|_{\bot}^2)^{\frac{1}{2}+i\nu}(2\cos^2(\phi_x)-1)}{(|x_1-z_0|_{\bot}^2)^{\frac{1}{2}+i\nu}|x_3-z_0|_{\bot}^2)^{\frac{1}{2}+i\nu}}
\frac{(|y_{13}|_{\bot}^2)^{\frac{1}{2}-i\nu}(2\cos^2(\phi_y)-1)}{(|y_1-z_0|_{\bot}^2)^{\frac{1}{2}-i\nu}|y_3-z_0|_{\bot}^2)^{\frac{1}{2}-i\nu}},
\label{corr int z0}\end{gather}
where \(\phi_x\) is the angle between the vectors \(z_{0}-x_{1\bot}\) and \(z_{0}-x_{3\bot}\),  \(\phi_y\) is the angle between the vectors \(z_{0}-y_{1\bot}\) and \(z_0-y_{3\bot}\).

The last integration can be done directly in the limit \(x_{13}, \ y_{13} \rightarrow 0\).  The calculations are given in Appendix \ref{Integrals}. Finally we get:

\begin{gather}
\langle S_{gl+}^{2+\omega_1}(x_{1\bot},x_{3\bot})S_{gl-}^{2+\omega_2}(y_{1\bot},y_{3\bot})\rangle \xrightarrow[x_{{}_{13\bot}}, \ y_{{}_{13\bot}} \rightarrow 0]\notag\\
\rightarrow-i\frac{N^2g^4}{4\pi^3}\delta(\omega_1-\omega_2)\int d \nu (\Delta_{\bot}^2)^{\aleph(\nu)-\omega}B(-\omega,\omega-\aleph(\nu))\frac{1-e^{i\pi(2\aleph(\nu)-\omega)}}{\sin \pi \aleph(\nu)}\frac{\nu^2}{(\frac{1}{4}+\nu^2)^2}\cdot\notag\\
\frac{1}{(|x_{13}|_{\bot}^2|y_{13}|_{\bot}^2)^{1+\frac{\aleph(\nu)}{2}}}\left(\frac{(|x_{13}|_{\bot}^2)^{\frac{1}{2}+i\nu}(|y_{13}|_{\bot}^2)^{\frac{1}{2}+i\nu}}{(|x-y|_{\bot}^2)^{1+2i\nu}}G(\nu)+(\nu\rightarrow -\nu) \right),
\end{gather}
where
\begin{equation*}
 G(\nu)=-i\frac{4^{-1-2i\nu}\pi^3(i-2\nu)^2}{\Gamma^2(\frac{3}{2}-i\nu)\Gamma^2(1+i\nu)\sinh(2\pi\nu)}\,.
\end{equation*}
Now we can carry out the last integration over \(\nu\) as the pole contribution at \(\o=\aleph(\nu)\).  We pick here the first pole  \(\Psi\)-functions in \eqref{Aleph}  which corresponds to the operator with the lowest possible  twist=2. Note that we omitted from our contour of integration the singularity at \(\nu=-\frac{i}{2}\).
Finally, we arrive at the final result of our paper - the correlator of two light ray operators  representing the analytic continuation of twist two operators  to the Lorentz spin \(j=1+\o\), in the BFKL limit \(\o\to 0\) and \(\frac{g^2}{\o}\) - fixed:
\begin{gather}
\langle S_{+}^{2+\omega_1}(x_{1\bot},x_{3\bot})S_{-}^{2+\omega_2}(y_{1\bot},y_{3\bot})\rangle \xrightarrow[x_{{}_{13\bot}}, \ y_{{}_{13\bot}} \rightarrow 0]{ } \delta(\omega_1-\omega_2)\Upsilon(\tilde{\gamma}) \frac{(x_{13\bot}^2)^{\frac{\tilde{\gamma}}{2}-\frac{\omega}{2}}(y_{13\bot}^2)^{\frac{\tilde{\gamma}}{2}-\frac{\omega}{2}}}{((x-y)_{\bot}^2)^{2+\tilde{\gamma}}},
\end{gather}
where  \(\Upsilon\) is given by
\begin{gather}
\Upsilon(\tilde{\gamma})=-N^2g^4\frac{2^{-1-2\tilde{\gamma}}\pi}{\tilde{\gamma}^2\Gamma^2(1-\frac{\tilde{\gamma}}{2})\Gamma^2(\frac{1}{2}+\frac{\tilde{\gamma}}{2})\sin(\pi\tilde{\gamma})\hat{\aleph}'(\tilde{\gamma})}
\end{gather}
and  \(\tilde{\gamma}=-1+2i\nu\) is the solution of \(\omega=\hat \aleph(\tilde{\gamma})\) , where \(\hat\aleph(\tilde{\gamma})=\aleph(-i\frac{\tilde{\gamma}+1}{2})\) and \(\aleph(\nu)\) is given by \ref{Aleph}.
Finally let us introduce the new quantity \(\gamma=\tilde{\gamma}-\omega\) which is the anomalous dimension in NLO BFKL. It satisfy the following equation:
\begin{gather}
\omega=\hat\aleph(\gamma+\omega)=\hat\aleph(\gamma)+\hat\aleph'(\gamma)\hat\aleph(\gamma)+o(g^4)\,.\label{NLOanomdim}
\end{gather}
This anomalous dimension \(\gamma\) is in the full correspondence with \cite{Kotikov:2002ab}. The correlator in terms of \(\gamma\) reads as follows:
\begin{gather}
\langle S_{+}^{2+\omega_1}(x_{1\bot},x_{3\bot})S_{-}^{2+\omega_2}(y_{1\bot},y_{3\bot})\rangle\xrightarrow[x_{{}_{13\bot}}, \ y_{{}_{13\bot}} \rightarrow 0]{ } \delta(\omega_1-\omega_2)\Upsilon(\gamma+\omega) \frac{(x_{13\bot}^2)^{\frac{\gamma}{2}}(y_{13\bot}^2)^{\frac{\gamma}{2}}}{((x-y)_{\bot}^2)^{2+\gamma+\o}}\,.
\label{final result}\end{gather}
Note that this formula correctly reproduces the tensor structure of the correlator corresponding  of local twist-2 operators  \eqref{LocOper} restricted on 2-dimensional orthogonal space and analytically continued to \(j\to 1+\o\). Indeed, the regularized operators enter with a multiplier \(\Lambda^{\gamma}\) where \(\Lambda\) is a scheme-dependent cutoff. We use the point-splitting regularization in the orthogonal direction for our light-ray operators and hence the cutoffs are defined as \(\Lambda_x=\frac{1}{|x_{13\bot}|}\) and \(\Lambda_y=\frac{1}{|y_{13\bot}|}\). Now if we redefine light ray operators as \(\Lambda_x^\gamma\breve{S}_{+}^{2+\omega_1}(x_{1\bot})\rightarrow \breve{S}_{+}^{2+\omega_1}(x_{1\bot})\) , \(\Lambda_y^\gamma\breve{S}_{+}^{2+\omega_1}(y_{1\bot})\rightarrow \breve{S}_{+}^{2+\omega_2}(y_{1\bot})\)  the correlation function acquires a standard form:
\begin{gather}
\langle \breve{S}_{+}^{2+\omega_1}(x_\bot)\breve{S}_{-}^{2+\omega_2}(y_\bot)= \delta(\omega_1-\omega_2) \frac{\Upsilon(\gamma+\omega)}{((x-y)_{\bot}^2)^{2+\gamma+\o}}\,.
\end{gather}
In the leading order of perturbation theory, when \(\frac{g^2}{\omega}\rightarrow 0\), the coefficient \(\Upsilon(\gamma+\o)\) reads as follows:
\begin{gather}
\Upsilon(-8{g^2}/{\omega})=\frac{\omega N^2}{\pi 2^7}
\end{gather}
and our BFKL result \eqref{final result} reduces  to \begin{gather}
\langle \breve{S}_{+}^{2+\omega_1}(x_\bot)\breve{S}_{-}^{2+\omega_2}(y_\bot)\rangle=\delta(\omega_1-\omega_2)\frac{\omega N^2}{\pi 2^7}  \frac{1}{((x-y)_{\bot}^2)^{2+\o}}\,.
\label{final result g=0}\end{gather}

\section{Conclusion}
In this paper we have generalized the twist-2 operator for the case of principal series representation in terms of a nonlocal light ray operator. Then we have calculated the correlation function between two such operators in the BFKL limit. The  correlator takes the form expected from conformal invariance, with the same anomalous dimension as predicted in  \cite{Kotikov:2002ab}.
One might ask why this predictable result could be interesting. Here is our motivation.

First of all, the method of \cite{Kotikov:2002ab} is rather indirect and is based on the the comparison with the Bjorken scaling for the scattering amplitudes. It was suggested there that an analytic continuation of anomalous dimensions
of local twist-2 operators gives the anomalous dimension of some non-local gluon operator
$F_{-i}\nabla^{\omega-1}F_-^{~i}$.
This method, however, does not tell us the explicit form of this operator and in this paper we demonstrated that $F_{-i}\nabla^{\omega-1}F_-^{~i}$ is actually a light-ray operator ($j\equiv \omega+1$):
\begin{eqnarray}
\mathcal{F}_j(x_\bot)
=\int_0^\infty \! dL_+ ~L_+^{1-j}\!\int\! dx_+~\tr F_-^{\,\,\,i}(x_+ n_-+x_\bot)[x,x+L_+n_-] F_{-i}\left((L_++x_+)n_-+x_\bot\right)
\nonumber\\
\label{ef}
\end{eqnarray}
with the anomalous dimension of this operator $\gamma(j;g^2)$ being an analytic continuation of the anomalous dimension (\ref{NLOanomdim}) of local twist two operators. The correlator we  calculate  is a physical quantity well adopted to the study of CFT.

Secondly, we hope to generalize this result to the case of three-point correlators (and the corresponding structure functions) of twist-2 operator in the BFKL limit.  An important basic ingredient for it -- the so called 3-pomeron transition vertex -- is already present in the literature \cite{Bartels:1993ih,Bartels:1994jj,Korchemsky:1997fy}. However, this vertex is a purely 2-dimensional object (in the  2d space orthogonal to the light-cone directions). In a work in progress, we are trying to understand how to adopt it in the context of  4-dimensional 3-point correlators and our experience  with the 2-point correlator
from the current paper,  and in particular its explicitly calculated normalization, can serve as a  valuable material.

In other words, our work can be considered as the first step in the construction of the conformal bootstrap ingredients for the 4-dimensional OPE in the BFKL approximation in N=4 SYM theory. This could provide a valuable information on the general structure of the OPE in this model beyond the perturbation theory.
An additional interest for it is the fact that in the leading BFKL order, the same gluon graphs dominate both the planar QCD and the planar N=4 SYM. This bootstrap program could  thus provide an interesting point of view to the hadron high energy scattering.

 And finally, let us stress again that our generalization of twist-2 operators based  on principal series representation with continuous spin \(j\) allows us to circumvent a subtle question of analytic continuation in \(j\).  The well-known principal of maximal transcendentality, which often serves as a mnemonic prescription for such  analytic continuation, notably in the perturbative expansion based on integrability \cite{Bajnok:2008qj,Lukowski:2009ce}, might naturally emerge in the framework of the extension of N=4 SYM physical space  to the principal series of \(PSU(2,2|4)\) or its subgroups. It is tempting to suggest that the principal series representation, in terms of nonlocal objects generalising local operators,  might  fix at once the analytic continuation for all such observables.

\section*{Acknowledgments}
We thank R.Janik, J.Penedones and A.Sever for the valuable discussions. We are especially grateful to G.Korchemsky for his contributions on the early stage of this work and  for generously sharing with us his knowledge on the subject.    E.S. thanks Perimeter Institute in Waterloo Canada and  Kavli IPMU in Tokyo for hospitality and UNIFY grant for support. V.K. and E.S. thank  e ANR grants StrongInt (BLANCSIMI- 4-2011) and GATIS network for the support.
 V.K. and E.S. also thank Yukawa Institute for Theoretical Physicsin Kyoto where the final part of the work was done, for the hospitality. The work of Ian Balitsky was supported by DOE grant DE-AC05-06OR23177.

 \section*{Appendices}

\appendix

\section{Notations}\label{AppNotations}

In this section we set our notations.
The lagrangian of N=4 SYM with the \(SU(N_c)\) gauge group has the following form:

\begin{gather}
\mathfrak{L}=\Tr\left\{-\frac{1}{2}F_{\mu\nu}F^{\mu\nu}+\frac{1}{2}(D_\mu\phi^{AB})(D^\mu\bar{\phi}_{AB}) +\frac{1}{8}g^2[\phi^{AB},\phi^{CD}][\bar{\phi}_{AB},\bar{\phi}_{CD}]+\right. \notag\\
\left.+2i\bar{\lambda}_{\dot{\alpha}A}\sigma_{\mu}^{\dot{\alpha}\beta}D^\mu\lambda_\beta^A-\sqrt{2}g\lambda^{\alpha A}[\bar{\phi}_{AB},\lambda^B_\alpha]+\sqrt{2}g\bar{\lambda}_{\dot{\alpha}A}[\phi^{AB},\bar{\lambda}_B^{\dot{\alpha}}]\right\},
\end{gather}
where field strength \(F_{\mu\nu}=\partial_\mu A_\nu-\partial_\nu A_\mu-ig[A_\mu,A_\nu]\) and covariant derivative \(D_\mu=\partial_\mu -ig[A_\mu,...]\).Notice that we work with Minkowski signature \((+,-,-,-)\) and all fields are taken in the adjoint representation of \(SU(N_c)\). \(SO(6)\)-multiplet with scalars \(\phi^a, a\in\{1\div6\}\) can be grouped into the antisymmetric tensor \(\phi^{AB}\),\(A,B\in\{1\div4\}\):
\begin{equation}
\phi^{AB}=\frac{1}{\sqrt{2}}\Sigma^{a AB}\phi^a, \ \ \ \ \bar{\phi}_{AB}=\frac{1}{\sqrt{2}}\bar{\Sigma}^a_{AB}\phi^a=(\phi^{AB})^*,
\end{equation}
using Dirac matrices in 6-d Euclidian space:
\begin{gather*}
\Sigma^{a AB}=(\eta_{1AB},\eta_{2AB},\eta_{3AB},i\bar{\eta}_{1AB},i\bar{\eta}_{2AB},i\bar{\eta}_{3AB}),\\
\bar{\Sigma}^a_{AB}=(\eta_{1AB},\eta_{2AB},\eta_{3AB},-i\bar{\eta}_{1AB},-i\bar{\eta}_{2AB},-i\bar{\eta}_{3AB}),
\end{gather*}
and 't Hooft symbols:
\begin{gather*}
\eta_{iAB}=\epsilon_{iAB}+\delta_{iA}\delta_{4B}-\delta_{iB}\delta_{4A},\\
\bar{\eta}_{iAB}=\epsilon_{iAB}-\delta_{iA}\delta_{4B}+\delta_{iB}\delta_{4A},
\end{gather*}

\begin{equation}
\eta_1=\begin{pmatrix}
0 & 0 & 0 & 1\\
0& 0 & 1 & 0\\
0& -1 & 0 & 0\\
-1 & 0&0&0\\
\end{pmatrix},
\eta_2=\begin{pmatrix}
0 & 0 & -1 & 0\\
0& 0 & 0 & 1\\
1& 0 & 0 & 0\\
0 & -1& 0& 0\\
\end{pmatrix},
\eta_3=\begin{pmatrix}
0 & 1 & 0 & 0\\
-1& 0 & 0 & 0\\
0& 0 & 0 & 1\\
0 & 0& -1& 0\\
\end{pmatrix},
\end{equation}

\begin{equation}
i\bar{\eta}_1=\begin{pmatrix}
0 & 0 & 0 & -i\\
0& 0 & i & 0\\
0& -i & 0 & 0\\
i & 0&0&0\\
\end{pmatrix},
i\bar{\eta}_2=\begin{pmatrix}
0 & 0 & -i & 0\\
0& 0 & 1 & -i\\
i& 0 & 0 & 0\\
0 & i& 0& 0\\
\end{pmatrix},
i\bar{\eta}_3=\begin{pmatrix}
0 & i & 0 & 0\\
-i& 0 & 0 & 0\\
0& 0 & 0 & -i\\
0 & 0& i& 0\\
\end{pmatrix}.
\end{equation}
Explicit formula for scalars reads as follows

\begin{gather*}
[\phi^{AB}]=\frac{1}{\sqrt{2}}(\phi^1 \eta_{1AB}+\phi^2\eta_{2AB}+\phi^3\eta_{3AB}+\phi^4i\bar{\eta}_{1AB}+\phi^5i\bar{\eta}_{2AB}+\phi^6i\bar{\eta}_{3AB})=\\
=\frac{1}{\sqrt{2}}\begin{pmatrix}
0 & \phi^3+i\phi^6 & -\phi^2-i\phi^5 & \phi^1-i\phi^4\\
-\phi^3-i\phi^6& 0 & \phi^1+i\phi^4 & \phi^2-i\phi^5\\
\phi^2+i\phi^5& -\phi^1-i\phi^4 & 0 & \phi^3-i\phi^6\\
-\phi^1+i\phi^4 & -\phi^2+i\phi^5& -\phi^3+i\phi^6& 0\\
\end{pmatrix}
=\begin{pmatrix}
0 & Z & -Y & \bar{X}\\
-Z& 0 & X & \bar{Y}\\
Y& -X & 0 & \bar{Z}\\
-\bar{X} & -\bar{Y}& -\bar{Z}& 0\\
\end{pmatrix}.
\end{gather*}
Fermions are realized as a two-component Weyl spinors \(\lambda_\alpha^A\) with conjugated \(\bar{\lambda}_{\dot{\alpha}A}\). Spinor index \(\alpha\in\{1,2\}\)and \(A\in\{1\div4\}\) is a \(SU(4)\) index. Due to supersymmetry one can fix just the propagator of scalars and get the normalization for fermions and gauge fields acting by supercharges. In this article we set the normalization for free propagators as follows:

\begin{gather}
\langle Z(x)^a_b \bar{Z}(y)^c_d\rangle_0=\mathcal{N}(\delta ^a_d\delta ^c_b-\frac{1}{N_c}\delta ^a_b\delta ^c_d)\frac{1}{(x-y)^2},\ \ \text{and the same for}\  X \text{and}\  Y,\\
\langle \lambda_{\alpha}^A(x)^a_b \bar{\lambda}_{\dot{\beta} B}(y)^c_d \rangle_0=i\mathcal{N}\delta^A_B(\delta ^a_d\delta ^c_b-\frac{1}{N_c}\delta ^a_b\delta ^c_d)\bar{\sigma}^\mu_{\alpha \dot{\beta}}\frac{\partial}{\partial x^\mu}\frac{1}{(x-y)^2},\\
\langle A_\mu(x)^a_b A_\nu(y)^c_d \rangle_0 =-\mathcal{N}(\delta ^a_d\delta ^c_b-\frac{1}{N_c}\delta ^a_b\delta ^c_d)\frac{g_{\mu \nu}}{(x-y)^2}.
\end{gather}
where \(\mathcal{N}=-\frac{1}{8\pi^2}\), \(\{\sigma^\mu\}=\{1,\mathbf{\sigma}\}\) and \(\{\bar{\sigma}^\mu\}=\{1,-\mathbf{\sigma}\}\) with ordinary Pauli matrices \(\mathbf{\sigma}\).
Throughout the text we use the basis \(\{n_+,n_-,e_{1\bot},e_{2\bot}\}\) with two  light-like vectors \(n_+^\mu=\{\frac{1}{\sqrt{2}},0,0,\frac{1}{\sqrt{2}}\},\ \ n_-^\mu=\{\frac{1}{\sqrt{2}},0,0,-\frac{1}{\sqrt{2}}\}\) normalized as \((n_- n_+)=1\) and two orthogonal vectors \(e_{1\bot},e_{2\bot}\) , which span 2-d plane \(\{\bot\}\) orthogonal to \(\{n_+,n_-\}\). The vector \(x\) reads in this basis  as      \(x=x_- n_+ +x_+n_-+x_\bot\),  with its square equal to \(x^2=2x_+x_--x_\bot^2\).

\par\medskip
\paragraph{Field content of twist-2 operators}
All twist-2 operators, which were discussed in this paper, are constructed from the set of elementary fields \(X=\{F^{\ \ \mu}_{+ \bot}, \lambda^A_{+\alpha},\bar{\lambda}^{\dot{\alpha}}_{+A},\phi^{AB}\}.\)  Twist 2 is the minimal possible twist (defined as bare dimension minus spin). Gluon field \(F^{\ \ \mu}_{+\bot}\) is obtained by projection of one of the indices of the field strength tensor \(F^{\mu\nu}\) on \(n^+\) direction where as  the second index is automatically restricted to the transverse plane with the metric \(g^\bot_{\mu\nu}=g_{\mu\nu}-n_{+\mu} n_{-\nu}-n_{+\nu} n_{-\mu}\). Weyl spinors \(\lambda_{+\alpha}\) and \(\bar{\lambda}_+^{\dot{\alpha}}\) correspond to the states with definite helicity \(1,-1\), respectively and they are parameterized as \(\lambda_{+\alpha}=\frac{1}{2}\bar{\sigma}^-_{\alpha\dot{\beta}}\sigma^{+\dot{\beta}\gamma}\lambda_\gamma\) and \(\bar{\lambda}_+^{\dot{\alpha}}=\frac{1}{2}\sigma^{-\dot{\alpha}\beta}\bar{\sigma}^+_{\beta \dot{\gamma}}\bar{\lambda}^{\dot{\gamma}}\).

\section{Explanation of the coordinate dependence of the cut-off ratio \eqref{cutoff} using NLO Impact factor }\label{CutOFFfromNLO}

\begin{figure}[H]
\center{\includegraphics[scale=1.0]{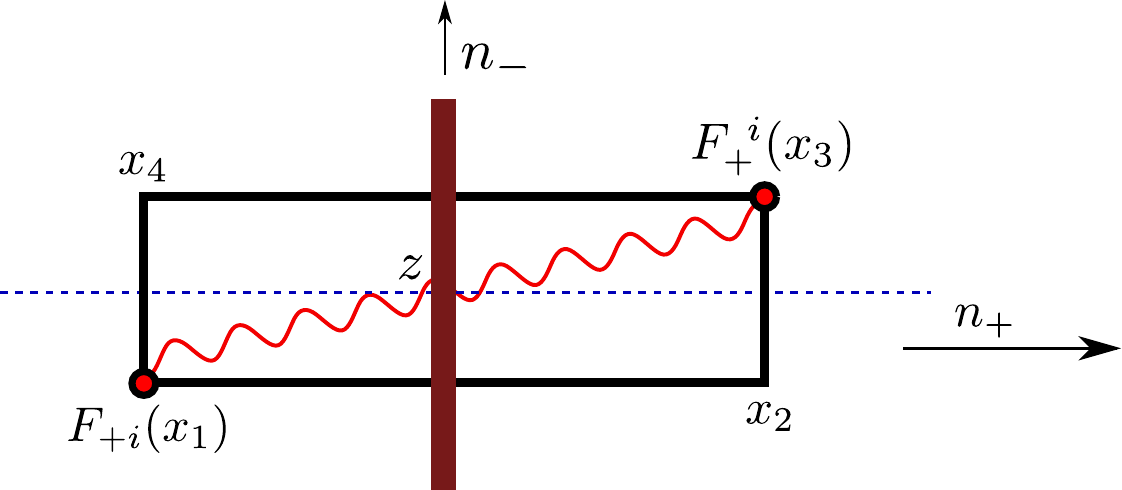}}
\caption{ImpactFactorLO  }
\label{ris:ImpactFactorLO}
\end{figure}

\begin{figure}[H]
\center{\includegraphics[scale=1.0]{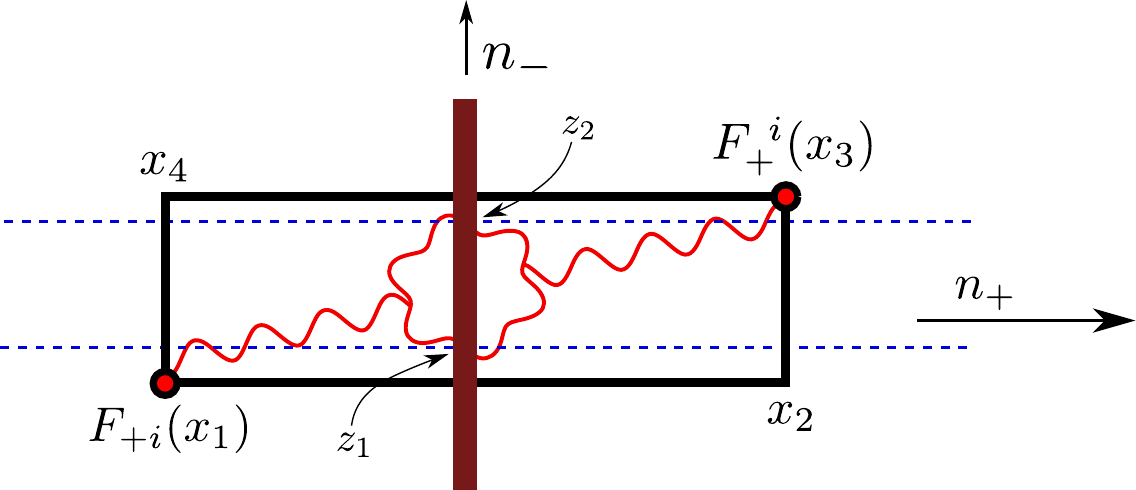}}
\caption{ImpactFactorNLO  }
\label{ris:ImpactFactorNLO}
\end{figure}

In principle, in the context of high energy scattering, the cutoffs $\sigma$ in Eq. (4.3) should be obtained from  the NLO impact factor for Wilson frame.
In accordance with general logic of high-energy OPE we factorize any correlation function into a product
of the ``probe'' impact factor, the ``target'' impact factor, and the amplitude of scattering of two (conformal) dipoles.
The ``rapidity divide'' between the impact factor and the dipole-dipole scattering is determined from two conditions:
(i) the properly defined impact factor should not scale with the energy, so that all the energy dependence is contained in the dipole-dipole scattering,
and (ii) the impact factor should be M\"obius invariant.
The calculation of the NLO impact factor for frames is beyond the scope of present paper where we limit ourselves only to the LO impact factor, with a typical Feynman graph  given in Fig.\ref{ris:ImpactFactorLO} (but take into account the NLO dimension!);  however, it is instructive to consider a typical  Feynman graph in NLO   to read off the cutoff dependence on the shape of the configuration of frames.  A typical Feynman diagram for the NLO impact factor is shown in Fig.\ref{ris:ImpactFactorNLO}
and the result is proportional to \cite{Balitsky:2010jf}

\begin{eqnarray}
g^2\int d^2z_1d^2z_2
\int_0^\infty dp_{1-} e^{i \frac{p_{1-}}{ 2}{\cal Z}_1}   \int_0^\infty{dp_{2-}\over p_{2-}}~e^{i \frac{p_{2-}}{ 2}{\cal Z}_2}
+(z_1\leftrightarrow z_2),   \label{fla1}
\end{eqnarray}
where  ${\cal Z}_i\equiv \frac{(x_1-z_i)_\perp^2}{ x_{1-}}-
\frac{(x_3-z_i)_\perp^2}{ x_{2-}}$. The integral over $\alpha_{2}$ in the Eq. (\ref{fla1}) diverges. This divergence reflects the fact that the eq.(\ref{fla1})  is not exactly the NLO impact factor since we must subtract from it the
matrix element of the leading-order contribution, the graph in Fig.\ref{ris:ImpactFactorLO}, which is proportional to
\begin{eqnarray}
&&\hspace{-1mm}
g^2\!\int\! d^2z_1\!\int_0^\infty\!dp_{1-}~e^{i{p_{1-}\over 2}{\cal Z}_1}   \!\int_0^{\sigma_-}\!{dp_{2-}\over p_{2-}},
\label{fla2}
\end{eqnarray}
where the integral over $p_{2-}$ is restricted by the ``rigid cutoff'' (\ref{cutoff}). The difference of these two expressions
gives the typical logarithmic term in the NLO impact factor in the form
\begin{gather}
g^2\int d^2z_1d^2z_2
\int_0^\infty dp_{1-}~e^{i{p_{1-}\over 2}{\cal Z}_1}  \Big( \!\int_0^\infty\!{dp_{2-}\over p_{2-}}~e^{i {p_{1-}\over 2}{\cal Z}_2}
-\int_0^{\sigma_-}{dp_{2-}\over p_{2-}}\Big)+(z_1\leftrightarrow z_2)=
\notag\\
={1\over {\cal Z}_1^2}\ln\sigma {\cal Z}_2~+~(z_1\leftrightarrow z_2).
\label{fla3}
\end{gather}
The logarithmic contribution is obviously not conformally invariant.  As explained in \cite{Balitsky:2010jf} the reason is that while formally light-like Wilson
lines are M\"obius invariant, the rigid cutoff (\ref{cutoff}) violates the invariance. Since the  conformally invariant cutoff for rapidity divergence
of Wilson lines is not known (it may even not exist) we proceed with the rigid cutoff (\ref{cutoff}) but pay the price of correcting
the ``rigid-cutoff'' dipoles by counterterms restoring the conformal invariance order-by-order in perturbation theory. In the NLO approximation
such ``composite conformal dipole'' has the form
\begin{gather}
\Ub(z_1,z_2)^{\rm conf}=\Ub(z_1,z_2)+\notag\\
+{g^2\over \pi}\!\int\! d^2 z_3~{z_{12}^2\over z_{13}^2z_{23}^2}
\left[\Ub(z_1,z_3)+
\mathcal{\Ub}(z_3,z_2)-\mathcal{\Ub}(z_1,z_2)\right]
 \ln {az_{12}^2\over z_{13}^2z_{23}^2}
\label{confodipole}
\end{gather}    is the ``composite dipole'' with the conformal longitudinal cutoff in the next-to-leading order and $a$ is an arbitrary dimensional constant.
The arbitrary dimensional constant $a$ should be chosen in such a way that the impact factor \eqref{fla1} does not change with length of the frame.
 It is convenient to choose the rapidity-dependent constant
$a\rightarrow ae^{-2\eta}$ so that the
$[{\rm Tr}\{\hat{U}^\sigma_{z_1}\hat{U}^{\dagger\sigma}_{z_2}\}\big]_a^{\rm conf}$
does not depend on $\eta=\ln\sigma_-$ and all the rapidity dependence is encoded into $a$-dependence:
\begin{gather}
\Ub(z_1,z_2)^{\rm conf}=\Ub(z_1,z_2)+\notag\\
+{g^2\over \pi}\!\int\! d^2 z_3~{z_{12}^2\over z_{13}^2z_{23}^2}
\left[\Ub(z_1,z_3)+
\mathcal{\Ub}(z_3,z_2)-\mathcal{\Ub}(z_1,z_2)\right]
\ln {2az_{12}^2\over \sigma_+^2 z_{13}^2z_{23}^2}~+~O(\alpha_s^2).
\label{confodipola}
\end{gather}
We need to choose the new ``rapidity cutoff'' $a$ in such a way that all the energy dependence is included into the matrix element(s) of
Wilson-line operators so that the impact factor does not depend on energy (i.e. it should not scale with the  length of frame.

Also, the NLO impact factor should be M\"obius invariant. These two requirements fix the cutoff in the form
$a_0={ 2x_{1-}x_{3-} \over (x-y)^2}$ and we obtain that the  typical logarithmic term in the NLO impact factor is proportional to
\begin{eqnarray}
&&\hspace{-1mm}
{1\over {\cal Z}_1^2}\Big[\ln{-x_{1-} x_{3-} z_{12}^2\over x_{13_\perp}^2 (x_{1_\perp}-z_2)^2z_{12}^2}
{\cal Z}_2^2+2C\Big]~+~(x_1\leftrightarrow x_3)~+~(z_1\leftrightarrow z_2).
\label{fla33}
\end{eqnarray}
Thus, the   ``new rapidity cutoff'' for the upper Wilson frame is $\sigma_-=\frac{ 2x_{1-}x_{3-} }{ x_{13_\perp}^2}$ (for simplicity, we use the same notation
$\sigma$ since the meaning of $a_0$ is essentially the rapidity cutoff for the conformal dipole (\ref{confodipole})). Similarly,
for the lower Wilson frame the cutoff is $\sigma_-~=~{ 2y_{1+}y_{3+} \over y_{13_\perp}^2}$ so we get $\sigma_+\sigma_-=r_1=r_2$ at large longitudinal $x,y$.

\section{ Calculation of the integral in \eqref{corr int z0} }\label{Integrals}

To carry out the integration over \(z_0\) in the integral
\begin{gather}
A_{\mathbb{R}^2}=\int\limits_{\mathbb{R}^2} d^2z_0\frac{(|x_{13}|^2)^{\frac{1}{2}+i\nu}(2\cos^2(\phi_x)-1)}{(|x_1-z_0|^2)^{\frac{1}{2}+i\nu}|x_3-z_0|^2)^{\frac{1}{2}+i\nu}}
\frac{(|y_{13}|^2)^{\frac{1}{2}-i\nu}(2\cos^2(\phi_y)-1)}{(|y_1-z_0|^2)^{\frac{1}{2}-i\nu}|y_3-z_0|^2)^{\frac{1}{2}-i\nu}}
\end{gather}
  let us define two functions
\begin{gather}
A_{\Omega}=\int\limits_{\Omega} d^2z_0\frac{(|x_{13}|^2)^{\frac{1}{2}+i\nu}(2\cos^2(\phi_x)-1)}{(|x_1-z_0|^2)^{\frac{1}{2}+i\nu}|x_3-z_0|^2)^{\frac{1}{2}+i\nu}}
\frac{(|y_{13}|^2)^{\frac{1}{2}-i\nu}(2\cos^2(\phi_y)-1)}{(|y_1-z_0|^2)^{\frac{1}{2}-i\nu}|y_3-z_0|^2)^{\frac{1}{2}-i\nu}},
\end{gather}

\begin{gather}
B_{\Omega}=\int\limits_{\Omega} d^2z_0\frac{(|x_{13}|^2)^{\frac{1}{2}+i\nu}}{(|x_1-z_0|^2)^{\frac{1}{2}+i\nu}|x_3-z_0|^2)^{\frac{1}{2}+i\nu}}
\frac{(|y_{13}|^2)^{\frac{1}{2}-i\nu}}{(|y_1-z_0|^2)^{\frac{1}{2}-i\nu}|y_3-z_0|^2)^{\frac{1}{2}-i\nu}}
\end{gather}\
and divide the full \(\mathbb{R}^2\) space into three domains
\begin{eqnarray*}
(1)\quad &&\Omega_0=|x_1-z_0|,|x_3-z_0|>q_x \wedge |y_1-z_0|,|y_3-z_0|>q_y\\
(2)\quad &&\Omega_x=|x_1-z_0|,|x_3-z_0| <  q_x \\
(3)\quad &&\Omega_y=|y_1-z_0|,|y_3-z_0|<q_y \end{eqnarray*}
where
\begin{gather*}
 q_x=\sqrt{|x_{13}||x-y|}, \  q_y=\sqrt{|y_{13}||x-y|}
\end{gather*}      and calculate the  difference \(A_\Omega-B_\Omega\) for each of them.

In the case (1) we can expand \(\cos^2\approx 1+o(|x_{13}|,|y_{13}|)\), then \(2cos^2-1\rightarrow 1\). The  difference \(A_{\Omega_0}-B_{\Omega_0}\) disappears in this domain. Now let us elaborate the case (2) (the case (3) is absolutely similar). In this case we  integrate over \(z\) inside the circle centered at \(x_1\sim x_3\), with the radius \(q_x\):
\begin{gather}
A_{\Omega_x}-B_{\Omega_x}=\frac{(|x_{13}|^2)^{\frac{1}{2}+i\nu}(|y_{13}|^2)^{\frac{1}{2}-i\nu}}{(|x-y|^2)^{1-2i\nu}}\int\limits_{|z-x|<q_x}d^2z_0\frac{2\cos^2(\phi_x)-1-1}
{(|x_1-z_0|^2)^{\frac{1}{2}+i\nu}(|x_3-z_0|^2)^{\frac{1}{2}+i\nu}}(1+o(\frac{q_x}{|x-y|}))=\notag\\
=-2\frac{(|x_{13}|^2)^{\frac{1}{2}+i\nu}(|y_{13}|^2)^{\frac{1}{2}-i\nu}}{(|x-y|^2)^{1-2i\nu}}\int\limits_{\mathbb{R}^2}d^2z_0\frac{\sin^2(\phi_x)}
{(|x_1-z_0|^2)^{\frac{1}{2}+i\nu}(|x_3-z_0|^2)^{\frac{1}{2}+i\nu}}(1+o(\frac{x_{13}}{q_x})).
\end{gather}
The last integral can be calculated in elliptic coordinates
\begin{gather}
|x_1-z_0|=\frac{|x_{13}|}{2}(\sigma+\tau),\notag\\
|x_3-z_0|=\frac{|x_{13}|}{2}(\sigma-\tau),
\end{gather}
which gives:
\begin{gather}
\int\limits_{\mathbb{R}^2}d^2z\frac{\sin^2(\phi_x)}
{(|x_1-z_0|^2)^{\frac{1}{2}+i\nu}(|x_3-z_0|^2)^{\frac{1}{2}+i\nu}}= 2^{3+4i\nu}(|x_{13}|^2)^{-2i\nu}\int\limits_1^\infty d\sigma\int\limits_{-1}^1d\tau \frac{\sqrt{(\sigma^2-1)(1-\tau^2)}}{(\sigma^2-\tau^2)^{2+2i\nu}}=\notag\\
=-\pi 2^{-1+4i\nu}(|x_{13}|^2)^{-2i\nu}\frac{\Gamma(-\frac{1}{2}-i\nu)\Gamma(1+i\nu)}{\Gamma(1-i\nu)\Gamma(\frac{3}{2}+i \nu)},
\end{gather}
where we have used the formula:
\begin{gather}
\int\limits_1^\infty d\sigma\int\limits_{-1}^1d\tau \frac{\sqrt{(\sigma^2-1)(1-\tau^2)}}{(\sigma^2-\tau^2)^{2+2i\nu}}=\int\limits_1^\infty d\sigma
\sqrt{-1+\sigma^2}\frac{1}{2} \pi  \left(\sigma^2\right)^{-2-2 i \nu } {}_2F_1\left(\frac{1}{2},2+2 i \nu ,2,\frac{1}{\sigma^2}\right)=\notag\\
=-\frac{\pi  \Gamma\left(-\frac{1}{2}-i \nu \right) \Gamma(1+i \nu )}{16 \Gamma(1-i \nu ) \Gamma\left(\frac{3}{2}+i \nu \right)}\,.
\end{gather}
Finally we get :
\begin{gather}
\delta_x=A_{\Omega_x}-B_{\Omega_x}=\frac{(|x_{13}|^2)^{\frac{1}{2}-i\nu}(|y_{12}|^2)^{\frac{1}{2}-i\nu}}{(|x-y|^2)^{1-2i\nu}}
\pi 2^{4i\nu}\frac{\Gamma(-\frac{1}{2}-i\nu)\Gamma(1+i\nu)}{\Gamma(1-i\nu)\Gamma(\frac{3}{2}+i \nu)}.
\end{gather}
And similar expression for \(A_{\Omega_y}-B_{\Omega_y}\):
\begin{gather}
\delta_y=A_{\Omega_y}-B_{\Omega_y}=\frac{(x_{13}|^2)^{\frac{1}{2}+i\nu}(|y_{13}|^2)^{\frac{1}{2}+i\nu}}{(|x-y|^2)^{1+2i\nu}}
\pi 2^{-4i\nu}\frac{\Gamma(-\frac{1}{2}+i\nu)\Gamma(1-i\nu)}{\Gamma(1+i\nu)\Gamma(\frac{3}{2}-i \nu)}.
\end{gather}
Expression for   \(B_{\mathbb{R}^2}\) (when \(\Omega=\mathbb{R}^2\)) in the limit \(x_{13}, \ y_{13} \rightarrow 0\) reads as follows:
\begin{gather}
B_{\mathbb{R}^2}=\int\limits_{\mathbb{R}^2} d^2z_0
\frac{(|x_{13}|^2)^{\frac{1}{2}+i\nu}}{(|x_1-z_0|^2)^{\frac{1}{2}+i\nu}(|x_3-z_0|^2)^{\frac{1}{2}+i\nu}}
\frac{(|y_{13}|^2)^{\frac{1}{2}-i\nu}}{(|y_1-z_0|^2)^{\frac{1}{2}-i\nu}(|y_3-z_0|^2)^{\frac{1}{2}-i\nu}}=\notag\\
=\frac{(|x_{13}|^2)^{\frac{1}{2}+i\nu}(|y_{13}|^2)^{\frac{1}{2}+i\nu}}{(|x-y|^2)^{1+2i\nu}}F(\nu)+(\nu\rightarrow -\nu),
\end{gather}
where \(F(\nu)=\frac{\pi2^{-4i\nu}}{2i\nu}\frac{\Gamma(\frac{1}{2}+i\nu)\Gamma(-i\nu)}{\Gamma(\frac{1}{2}-i\nu)\Gamma(i\nu)}\).
Finally, collecting the individual terms we obtain
\begin{eqnarray*} A_{\mathbb{R}^2}=\frac{(|x_{13}|^2)^{\frac{1}{2}+i\nu}(|y_{13}|^2)^{\frac{1}{2}+i\nu}}{(|x-y|^2)^{1+2i\nu}}G(\nu)+(\nu\rightarrow -\nu),\end{eqnarray*}
where
\begin{equation*}
 G(\nu)=-i\frac{4^{-1-2i\nu}\pi^3(i-2\nu)^2}{\Gamma^2(\frac{3}{2}-i\nu)\Gamma^2(1+i\nu)\sinh(2\pi\nu)}\,.
\end{equation*}

\section{Two-point correlator of Wilson frames}
As was noticed before, the method of OPE over colour dipoles is quite general and can be applied to many different operators. In this section we give the expression for the case of pure Wilson frames (with no field insertion). Namely, such an operator for a frame stretched along \(n_+\) reads as follows:
\begin{gather}
S_{{}_{W.F.}+}^\omega(x_{1\bot},x_{3\bot})=\int\limits_{-\infty}^{\infty} d x_{1-} \int\limits_{x_{1-}}^{\infty}dx_{3-} (x_{3-}-x_{1-})^{-2-\omega}\ \tr[x_1,x_2]_{{}_\Box}.
\end{gather}
The operator constructed from a pure Wilson rectangle collapses to one when it is reduced to light-ray, but  it has a nontrivial correlation function when its transverse size is slightly different from zero.
The OPE expansion of frames over colour dipoles consists of  simply replacement of a finite frame by  an infinite dipole with a certain cutoff \(\s_+\):
\begin{gather}
\tr[x_1,x_3]_{{}_\Box}\rightarrow N(1-\Ub^{\sigma_+}(x_{1\bot},x_{3\bot})).
\end{gather}
This formula is an analogue of (\ref{FFpropagator}). The rest of calculation almost directly repeats the calculations for the regularized light ray operators of the main text and the result reads as follows:
\begin{gather}
\langle S_{{}_{W.F.}+}^{\omega_1}(x_{1\bot},x_{3\bot}) S_{{}_{W.F.}-}^{\omega_2}(y_{1\bot},y_{3\bot}) \rangle \sim \frac{g^4}{\omega}\frac{(x_{13\bot}^2)^{2+\frac{\gamma}{2}}(y_{13\bot}^2)^{2+\frac{\gamma}{2}}}{((x-y)_{\bot}^2)^{2+\gamma+\o}},\label{ramki otvet}
\end{gather}
 where \(\gamma\) is the anomalous dimension in the NLO BFKL given  by the  solution of (\ref{NLOanomdim}).
Let us stress that this result is in  correspondence with (\ref{final result}). Namely let us check the weak coupling regime \(\frac{g^2}{\omega}\rightarrow 0\). In this case we have :
\begin{gather}
\partial _{x_{1\bot}} \partial _{x_{3\bot}}\int\int (x_{3-}-x_{1-})^{-2-\omega}[x_1,x_3]_{{}_\Box}\simeq \frac{g_{{}_{YM}}^2}{\omega}\int\int (x_{3-}-x_{1-})^{-\omega}F(x_1)[x_1,x_3]_{{}_\Box}F(x_3).
\end{gather}
So it leads to the following correlator of two frames
\begin{gather}
\langle S_{+}^{2+\omega_1}S_{-}^{2+\omega_2}\rangle \sim
(\frac{\omega}{g_{{}_{YM}}^2})^2\partial _{x_{1\bot}} \partial _{x_{3\bot}} \partial _{y_{1\bot}} \partial _{y_{3\bot}}\langle S_{{}_{W.F.}+}^{\omega_1}(x_{1\bot},x_{3\bot}) S_{{}_{W.F.}-}^{\omega_2}(y_{1\bot},y_{3\bot}) \rangle,
\end{gather}
which coincides with (\ref{final result g=0}) and (\ref{ramki otvet}).

\printindex

\end{document}